\documentclass[10pt,twocolumn,journal]{IEEEtran}

\usepackage{amsmath}
\usepackage{amssymb}
\usepackage[dvips]{graphicx}
\usepackage{epsfig}

\newcommand{\E}{\mathbb{E}}
\newcommand{\en}{\mathcal{E}}

\newcommand{\C}{{\sf{C}}}

\newcommand{\bLambda}{\boldsymbol{\Lambda}}

\newcommand{\so}{\mathcal{S}_0}
\newcommand{\sinf}{\mathcal{S}_{\infty}}

\newcommand{\G}{\mathbf{G}}
\newcommand{\J}{\mathbf{J}}
\newcommand{\bpi}{\boldsymbol{\pi}}

\newcommand{\sps}{\text{sp}}
\newcommand{\on}{\text{ON}}
\newcommand{\avg}{\text{avg}}
\newcommand{\g}{\text{g}}
\newcommand{\cov}{\text{cov}}

\newcommand{\tmin}{\text{min}}

\newcommand{\tsnr}{{\text{\footnotesize{SNR}}}}

\newcommand{\ssnr}{\text{\scriptsize{SNR}}}

\newcommand{\tth}{\text{th}}

\newtheorem{theo}{Theorem}

\newtheorem{rem}{Remark}

\begin{document}

\title{Wireless Throughput and Energy Efficiency
with Random Arrivals and Statistical Queueing Constraints}

%

\author{Mustafa Ozmen and M. Cenk Gursoy \thanks{The authors are with the Department of Electrical Engineering and Computer Science,
Syracuse University, Syracuse, NY 13244. (e-mails:
mozmen@syr.edu, mcgursoy@syr.edu)}
\thanks{The material in this paper was presented in part at the 2012 IEEE International Symposium on Information Theory.}
}


%


\maketitle

\thispagestyle{empty}

\begin{abstract}
Throughput and energy efficiency in fading channels are studied in the presence of randomly arriving data and statistical queueing constraints. In particular, Markovian arrival models including discrete-time Markov, Markov fluid, and Markov-modulated Poisson sources are considered. Employing the effective bandwidth of time-varying sources and effective capacity of time-varying wireless transmissions, maximum average arrival rates in the presence of statistical queueing constraints are characterized. For the two-state (ON/OFF) source models, throughput is determined in closed-form as a function of the source statistics, channel characteristics, and quality of service (QoS) constraints. Throughput is further studied in certain asymptotic regimes. Furthermore, energy efficiency is analyzed by determining the minimum energy per bit and wideband slope in the low signal-to-noise ratio (SNR) regime. Overall, the impact of source characteristics, QoS requirements, and channel fading correlations on the throughput and energy efficiency of wireless systems is identified.
\end{abstract}

\section{Introduction} \label{sec:intro}

\subsection{Motivation}

Mobile data traffic has experienced unprecedented growth recently and is predicted to grow even further over the coming years. For instance,
it is projected that global mobile data traffic, which already grew 81 percent in 2013, will increase 11-fold between 2013 and 2018, reaching 15.9 exabytes per month by 2018 \cite{Cisco}. As one of the main driving forces behind this growth, wireless transmission of multimedia content has significantly increased in volume and is expected to be the dominant traffic in data communications. Indeed, mobile video traffic was already 53 percent of the traffic by the end of 2013 and is predicted to  increase 14-fold between 2013 and 2018, accounting for over two-thirds of the world's mobile data traffic by 2018 \cite{Cisco}.

This exponential growth in the flow of mobile data and multimedia content has significant implications on wireless networks. For one, wireless multimedia traffic requires certain quality-of-service (QoS) guarantees. For instance, in voice over IP (VoIP), multimedia streaming, interactive video, and online gaming applications, constraints on delay, packet loss, or buffer overflow probabilities need to be imposed  so that acceptable performance and quality levels can be met for the end-users.
Another consequence is heterogeneity in network traffic. Wireless networks now carry heterogeneous traffic in diverse environments, and successful design of networks, efficient use of resources, and effective QoS provisioning for multimedia communications critically depend on the appropriate choice of source traffic models. For instance, while voice traffic can be accurately modeled as an ON/OFF process with fixed-rate data arrivals in the ON state, data traffic can be bursty and video traffic, which exhibits correlations, can be modeled statistically using autoregressive, Markovian, or Markov-modulated processes \cite{survey-VBRvideotraffic}.

Finally, it is important to note that this increased traffic together with the given QoS requirements need to be supported by wireless systems equipped with only limited bandwidth and power resources. Especially, due to limited energy available for mobile units and rising energy costs and environmental concerns, energy efficiency in wireless communications is a key concern (see e.g., \cite{survey-Feng} and \cite{survey-Hasan}). Therefore, it is crucial to identify the fundamental performance limits (e.g., in terms of maximum achievable throughput and minimum energy per bit) in order to determine how to most effectively utilize the scarce resources. With this motivation, in this paper we investigate the throughput and energy efficiency in fading channels when data arrivals are in general random, and QoS constraints in the form of limitations on the asymptotic buffer overflow probabilities are imposed.



\subsection{Literature Overview} \label{subsec:literature}

Satisfying QoS requirements is critical for most communication networks, and how to satisfy QoS constraints for various source traffic models has been one of the key considerations in the networking literature. In particular, besides conventional queueing theory, network calculus has been introduced by Cruz in early 1990s as a theory to address the delay and other deterministic service guarantees in networks by dealing with queueing systems \cite{cruz_part1} -- \cite{cruz-qos}. Subsequently, Chang in \cite{chang} developed the stochastic version of the network calculus. More specifically, the theory of effective bandwidth of a
time-varying source has been formulated to identify the minimum
amount of transmission rate that is needed to satisfy the
statistical QoS requirements (see also \cite{changEB} -- \cite{Kelly}). This theory is based on the logarithmic moment generating function of the arrival process and is related to the large deviation principle. Moreover, statistical QoS constraints are imposed as limitations on buffer/delay violation probabilities. Effective bandwidths of various source models have been investigated extensively in the literature. For instance, Elwalid and Mitra studied the effective bandwidth of Markovian traffic sources (including Markov-modulated fluid and Markov-modulated Poisson sources) in \cite{elwalid} under constraints on the buffer overflow probability. It is shown that effective bandwidth is given by the maximum eigenvalue of a matrix derived from source parameters and service requirements.
In \cite{ebw}, effective bandwidth formulations were provided for multi-class Markov fluids as well as memoryless (Poisson) and discrete-time Markov sources. In \cite{costasweber}, the authors studied the effective bandwidths of general stationary sources and derived a first order approximation of the effective bandwidth in terms of the mean arrival rate and index of dispersion.

In wireless communications, the instantaneous channel capacity
varies randomly depending on the channel conditions. Hence, in
addition to the source characteristics, transmission rates for reliable
communication are also time-varying. In such cases, randomly time-varying servers can be considered in the queueing system model. Indeed, motivated by the wireless channel, Stolyar in \cite{Stolyar}, Venkataramanan and Lin in \cite{lin}, and Sadiq and de Veciana in \cite{Sadiq} employed tools from the theory of large deviations and investigated scheduling rules (e.g., MaxWeight, Exponential, and Radial Sum-Rate Monotonic scheduling) while controlling the large deviations of queues. Following another method, the time-varying channel capacity can be incorporated into the theory of effective bandwidth by regarding the channel service process as a time-varying source with negative rate and using the source multiplexing rule (\cite[Example 9.2.2]{Changbook}). Using a similar approach, as a dual concept to effective bandwidth, Wu and Negi defined in \cite{dapeng} the effective capacity, which describes the
maximum constant arrival rate that a given time-varying service
process can support while satisfying the statistical QoS requirements. Indeed, work in \cite{dapeng} revitalized the consideration of statistical queueing constraints in the context of wireless communications, and the effective
capacity of wireless transmissions has been investigated intensively in various settings
(see e.g., \cite{dw}--\cite{liu-energyeff}). For
instance,  Tang and Zhang in \cite{jia} considered the
effective capacity when both the receiver and transmitter know the
instantaneous channel gains, and derived the optimal power policy that maximizes the system throughput under QoS
constraints. Liu \emph{et al.} in \cite{liu} considered fixed-rate transmission schemes and analyzed the effective capacity and related resource requirements for Markov wireless channel models and Markov fluid sources. In \cite{aissa} and \cite{akin}, effective capacity of cognitive radio channels was studied. In \cite{gursoy-mimo}, multi-antenna communication in the presence of queueing limitations was investigated. Soret \emph{et al.} in \cite{Soret} addressed correlated Rayleigh fading channels and studied the effective capacity under different adaptive rate policies. In this study, performance in the presence of probabilistic delay constraints and variable rate sources was also analyzed by considering a Gaussian autoregressive source model. Energy efficiency in the presence of QoS limitations was addressed in \cite{gursoy-Twireless09} -- \cite{liu-energyeff}.


\subsection{Contributions}

We note that the studies on the effective capacity of wireless channels have primarily concentrated on constant arrival rates in the analysis of the throughput and energy efficiency\footnote{To the best of our knowledge, the two exceptions to this are references \cite{liu} and \cite{Soret} as also described in Section \ref{subsec:literature}. However, these studies have different modeling assumptions for the sources and/or wireless transmissions from what we have in this paper. For instance, in \cite{liu}, while ON/OFF Markov fluid arrivals are considered, wireless transmissions occur at fixed rates and wireless link is also modeled as a continuous-time Markov chain with ON and OFF states. In \cite{Soret}, a Gaussian autoregressive source is considered in the analysis. Additionally, our analysis, maximum average arrival rate expressions, and throughput characterizations in the low-$\theta$, high- and low-SNR regimes are novel contributions with respect to these prior studies. }. Departing from this approach, we in this paper explicitly take into account the randomness and burstiness of the source traffic. In particular, we address Markovian source models including discrete-time Markov, Markov fluid, and Markov modulated Poisson sources, and conduct a performance analysis. More specifically, our contributions can be listed as follows:
\begin{itemize}
\item A framework with which source randomness can be incorporated in the throughput analysis of wireless transmissions is provided.
\item For two-state (ON/OFF) source models, closed-form expressions are obtained for the maximum average arrival rate in terms of the source statistics, effective capacity of wireless transmissions, and the QoS exponent $\theta$, which quantifies how strict the QoS constraints are.
\item Throughput is characterized in the low-$\theta$ and high-SNR regimes.
\item An energy efficiency analysis is conducted and minimum energy per bit and wideband slope expressions are determined for both constant and random arrival models.
\item Via both analytical and numerical results, the impact of source randomness, fading correlations, and queueing constraints on the wireless throughput and energy efficiency is identified.
\end{itemize}

The remainder of this paper is organized as follows. In Section \ref{subsec:channelmodel}, we describe the channel model. Sections \ref{sec:energyefficiency} and \ref{subsec:effective-bandw-of-sources} contain the preliminaries regarding the statistical queueing constraints, effective bandwidth, and effective capacity. In Section \ref{sec:metrics}, we provide our characterizations of the throughput with Markovian source models by analyzing the maximum average arrival rates. We conduct an energy efficiency analysis in Section \ref{sec:EE} for both constant and Markovian arrivals. Finally, concluding remarks are given in Section \ref{sec:conclusion}. Proofs are relegated to the Appendix.


\section{System Model} \label{sec:chqueue}

\subsection{Channel Model} \label{subsec:channelmodel}
\begin{figure}
\begin{center}
\includegraphics[width=0.45\textwidth]{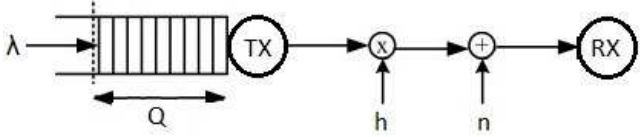}
\vspace{0.1cm}
\caption{System Model.} \label{fig:model}
\end{center}
\end{figure}

As depicted in Fig. \ref{fig:model}, we consider a point-to-point link with a single transmitter and single receiver. In this system, the data generated by the source is initially stored in a buffer at the transmitter before it is transmitted over a wireless channel. We consider a flat-fading channel between the transmitter and receiver, and assume a block-fading model with a block duration of $m$ symbols. Hence, fading varies independently from one block to another. On the other hand, we further assume that in each block duration of $m$ symbols, fading can be arbitrarily correlated. The channel input-output relation within each block can be expressed as
\begin{gather}
y_i = h_ix_i + n_i \text{ for } i = 1,2, \ldots ,m \label{eq:i-o}
\end{gather}
where $x_i$ and $y_i$ are the channel input and output, respectively. The average energy of the input is $\en$, i.e.,
\begin{gather}
\E\{|x_i|^2\} = \en.
\end{gather}
 $n_i$ denotes the zero-mean, circularly-symmetric, complex Gaussian noise with variance $\E\{|n_i|^2\} = N_0$. Hence, the signal-to-noise ratio is
\begin{gather}
\tsnr = \frac{\E\{|x|^2\}}{\E\{|n|^2\}} = \frac{\en}{N_0}. \label{eq:snr}
\end{gather}
Above in (\ref{eq:i-o}), $h_i$ denotes the fading coefficient. Fading coefficients are assumed to be identically distributed, and the fading distribution can be arbitrary with finite variance. 

While the ensuing analysis is applicable to a general class of fading distributions, we use a Gauss-Markov fading model in the numerical results and assume that the Gaussian fading coefficients in each block of $m$ symbols follow the correlation pattern $h_i = \rho h_{i-1} + w_i$ where $w_i$ is an independent, zero-mean Gaussian random variable with variance $\E\{|w_i|^2\} = (1-\rho^2) \sigma_h^2$, $\rho \in [0,1]$, and $\sigma_h^2$ is the common variance of the fading coefficients $\{h_i\}$. Note that when $\rho = 1$, we have full correlation, whereas $\rho = 0$ models the case of independent fading.

\subsection{Queueing Constraints} \label{sec:energyefficiency}
We assume that the data to be transmitted is generated from random sources and is first stored in a buffer before transmission. Statistical constraints are imposed on the queue length. In particular, we assume that the buffer violation/overflow probability satisfies
\begin{equation} \label{eq:theta}
\lim_{q \to \infty} \frac{\log \Pr\{Q \ge q\}}{q} = -\theta
\end{equation}
where $Q$ denotes the stationary queue length, and $\theta$ is the decay rate of the tail distribution of the queue length. The above limiting formula implies that for large $q$, we have
\begin{gather}
\Pr\{Q \ge q\} \approx e^{-\theta q}. \label{eq:overflowprob}
\end{gather}
Indeed, a closer approximation is \cite{dapeng}
\begin{gather}
\Pr\{Q \ge q\} \approx \varsigma e^{-\theta q} \label{eq:overflowprob-rev}
\end{gather}
where $\varsigma = \Pr\{Q > 0\}$ is the probability of non-empty buffer\footnote{Probability of non-empty buffer can be approximated from the ratio of average arrival rate to average service rate \cite{helmy-musavian}.}.
From (\ref{eq:overflowprob-rev}), we notice that, for a sufficiently large threshold, the buffer overflow probability should decay exponentially with rate controlled by the QoS exponent $\theta$. Note that as $\theta$ increases, stricter queueing or QoS constraints are imposed, while looser queueing constraints are implied by smaller values of $\theta$.  Conversely, for a given buffer threshold $q$ and overflow probability limit $\epsilon = \Pr\{Q \ge q\}$, the desired value of $\theta$ can be determined as
\begin{gather}
\theta = \frac{1}{q}\log_e \frac{\varsigma}{\epsilon}.
\end{gather}

In the given setting, the delay violation probability is also characterized to decay exponentially and is approximated by \cite{Du-Zhang}
\begin{gather}
\Pr\{D \ge d\} \approx \varsigma e^{-\theta a^*(\theta) d}  \label{eq:delayviolation}
\end{gather}
where $D$ is the queueing delay in the buffer at steady state, $d$ is the delay threshold, and $a^*(\theta)$ is the effective bandwidth of the arrival process, described below.

Next, we introduce the notions of effective bandwidth and effective capacity which we subsequently employ to formulate the wireless throughput in fading channels in the presence of random arrivals and statistical queueing constraints.

\subsubsection{Effective Bandwidth}
Effective bandwidth characterizes the minimum constant transmission (or service) rate required to support the given random data arrival process while the buffer overflow probability is limited or more explicitly the statistical queueing constraint described by (\ref{eq:theta}) is satisfied. Let $\{a(k), k=1,2,\ldots\}$ be a sequence of nonnegative random variables, describing the random arrival rates. Let also the time-accumulated arrival process be denoted by $A(t) = \sum_{k=1}^{t} a(k)$. Then, the effective bandwidth is given by the asymptotic logarithmic moment generating function of $A(t)$~\cite{chang}, i.e.,
\begin{gather}
a^*(\theta)= \lim_{ t \rightarrow \infty}\frac{1}{\theta t}\log \E \left\{e^{\theta A(t)}\right\}.
\end{gather}

In Section \ref{subsec:effective-bandw-of-sources}, we describe the effective bandwidth of different source arrival models in detail.

\subsubsection{Effective Capacity}
Effective capacity, as a dual concept to effective bandwidth, identifies the maximum constant arrival rate that can be supported by a given time-varying service process while satisfying \eqref{eq:theta}. Let $\{\nu[k], k=1,2,\ldots\}$ denote the discrete-time stationary and
ergodic stochastic service process and $S[t]\triangleq
\sum_{k=1}^{t} \nu[k]$ be the time-accumulated service process. Then, the effective capacity  is given by~\cite{dapeng}
\begin{equation}
C_E(\tsnr,\theta)=-\lim_{t\rightarrow\infty}\frac{1}{\theta
t}\log_e{\mathbb{E}\left\{e^{-\theta S[t]}\right\}}.
\end{equation}
Note that we have assumed that the fading coefficients $\{h_i\}$ change independently from one block of $m$ symbols to another. Under this assumption, effective capacity simplifies to
\begin{equation}\label{ec}
C_{E}(\tsnr,\theta)=-\frac{1}{\theta}\log_e\mathbb{E}\left\{e^{-\theta
\nu}\right\}
\end{equation}
where $\nu$ is the instantaneous service (or equivalently transmission) rate  in one block. If the channel input sequence $\{x_i\}$ is an independent and identically distributed (i.i.d.) sequence of Gaussian random variables with zero mean and variance $\en$, then the service rate
can be written as
\begin{gather}
\nu = \sum_{i = 1}^{m} \log_2(1 + \tsnr z_i)
\end{gather}
where we have defined $z_i = |h_i|^2$. Hence, the effective capacity in the units of bits/block is
\begin{align}
C_E(\tsnr,\theta) = -\frac{1}{\theta}\log_e\E \left\{e^{-\theta
\sum_{i = 1}^{m} \log_2(1 + \ssnr z_i)}\right\}. \label{eq:effcap}
\end{align}
%
\begin{rem}
In the special case of independent channel coefficients in each block and Rayleigh fading, we can express the effective capacity in closed-form as
\begin{align}
C_E(\tsnr, \theta)&\!=-\frac{m}{\theta}\log_e \!\! \left[\tsnr^{-\frac{\theta}{\log_e\!2}}e^{\frac{1}{\ssnr}} \,\Gamma\!\!\left(\!1\!-\!\frac{\theta}{\log_e\!2},\frac{1}{\ssnr}\right)\right] \label{eq:cesnrclosed}
\\
&\!=m\log_2(\tsnr)-\frac{m}{\theta\tsnr}-\frac{m}{\theta} \log_e\Gamma\!\!\left(\!1\!-\!\frac{\theta}{\log_e\!2},\frac{1}{\ssnr}\right) \label{eq:cesnrclosed}
\end{align}
where $\Gamma(s,w)=\int^{\infty}_w \tau^{s-1}e^{-\tau} d\tau$ is the upper incomplete gamma function.
\end{rem}

\subsection{Effective Bandwidths of Different Source Models} \label{subsec:effective-bandw-of-sources}

\subsubsection{Discrete-Time Markov Sources} \label{subsubsec:dMarkov}

In this subsection, we consider discrete-time Markov source models.  Assume that the transition probability matrix of the $n$-state irreducible and aperiodic Markov source process is denoted by $\J$, and $\lambda_i$ is the arrival rate in state $i$. Moreover, $\bLambda=diag\left\{\lambda_1,\lambda_2,\ldots,\lambda_n\right\}$ is the diagonal matrix of arrival rates. Then, the effective bandwidth of this discrete Markov source is given by \cite{Changbook}
\begin{align}
a(\theta)=\frac{1}{\theta}\log_e\left[\sps\left(e^{\theta \bLambda }\J\right)\right] \label{eq:dEBW}
\end{align}
where $\sps(\cdot)$ is the spectral radius of the input matrix. Note that the stationary distribution $\bpi$ can be found from the solution of
\begin{align}
\bpi\mathbf{1}&=1, \nonumber
\\
\bpi\J&=\bpi\label{eq:dstateprob}
\end{align}
where $\bpi=[\pi_1, \pi_2, \ldots ,\pi_n]$ and $\mathbf{1}=[1,\ldots,1]^T$.

In order to unveil the key relationships and tradeoffs, we consider a particularly simple two-state model.  We assume that data arrival is either in the ON or OFF state in each block duration of $m$ symbols. When the state is ON, $\lambda$ bits arrive (i.e., the arrival rate is $\lambda$ bits/block), while there are no arrivals in the OFF state. For this two-state model, the state transition probability matrix is given as
\begin{align}
\J=\left[
\begin{matrix}
 p_{11} & p_{12} \\
 p_{21} & p_{22}
\end{matrix}\right]. \label{eq:probmatrixtwostate}
\end{align}
Given the above transition matrix $\J$, the effective bandwidth for this ON-OFF Markov model can be derived as \cite{Changbook}
\begin{align}
a^*(\theta, \lambda) = \frac{1}{\theta} \log_e\!\!\left(\!\!\tfrac{p_{11}+p_{22} e^{\lambda\theta}+\sqrt{ (p_{11}+p_{22}e^{\lambda\theta})^2 - 4(p_{11}+p_{22}-1)e^{\lambda\theta} }  }{2}\right) \label{eq:2discreteEBW}
\end{align}
where $p_{11}$ denotes the probability of staying in the OFF state from one block to another. Similarly, $p_{22}$ denotes the probability of staying in the ON state. The probabilities of transitioning from one state to a different one are therefore denoted by $p_{21} = 1 - p_{22}$ and $p_{12} = 1 - p_{11}$. For these transition probabilities, we can easily see that the probability of the ON state in the steady state is
\begin{gather}
P_{\on} = \frac{1-p_{11}}{2-p_{11} - p_{22}}. \label{eq:P_ON}
\end{gather}
Therefore, the average arrival rate is
\begin{gather}
r_{\avg} = \lambda P_{\on} = \lambda \, \frac{1-p_{11}}{2-p_{11} - p_{22}} \label{eq:ravg_rPon}
\end{gather}
which is equal to the average departure rate when the queue is in steady state \cite{ChangZajic}.

\subsubsection{Markov Fluid Sources} \label{subsubsec:fMarkov}
In this subsection, we address Markov fluid sources where the source arrival process is modeled as a continuous-time Markov chain. Assume that $\G$ is the irreducible transition rate matrix of the Markov chain, $\lambda_i$ is the arrival rate in the $i^{\tth}$ state, and $\bLambda=diag\left\{\lambda_1,\lambda_2,\ldots,\lambda_n\right\}$. Then, the effective bandwidth of this source is given by \cite{elwalid}, \cite{ebw}
\begin{align}
a^*(\theta)=\mu\left(\bLambda+\frac{1}{\theta}\G\right) \label{eq:fluidebw}
\end{align}
where $\mu(\cdot)$ denotes the maximum real eigenvalue of the input matrix. We also note that the stationary distribution $\bpi$ of the continuous-time Markov chain can be found by solving
\begin{align}
\bpi\mathbf{1}&=1, \nonumber
\\
\bpi\G&=\mathbf{0} \label{eq:stateprob}
\end{align}
where $\bpi=[\pi_1, \pi_2, \ldots, \pi_n]$, $\mathbf{0}=[0, \ldots, 0]^T$ and $\mathbf{1}=[1, \ldots, 1]^T$.


In order to derive closed-form expressions in our analysis, we again consider two states (ON/OFF). When there is no arrival, the state is OFF. When the state is ON, the arrival rate is $\lambda$ bits/block. The transition rate matrix for a two-state Markov fluid is in the form of
\begin{align}
\G=\left[
\begin{matrix}
 -\alpha & \alpha \\
 \beta & -\beta
\end{matrix}\right], \label{eq:generating}
\end{align}
where $\alpha$ is the transition rate from OFF state to ON state whereas $\beta$ is the transition rate from ON state to OFF state. Using \eqref{eq:fluidebw}, we can express the effective bandwidth as
\begin{align}
a^*(\theta)=\frac{1}{2\theta}\left[\theta \lambda -(\alpha+\beta)+\sqrt{(\theta \lambda -(\alpha+\beta))^2+4\alpha\theta \lambda}  \right]. \label{eq:2fluidEBW}
\end{align}
The probability of ON state, $\pi_2$,  is required to define the average rate. Inserting the generator matrix $\G$ in (\ref{eq:generating}) into (\ref{eq:stateprob}), we obtain the ON state probability as
\begin{gather}
\pi_2 = P_{\on} = \frac{\alpha}{\alpha+\beta}. \label{eq:Pontwostate}
\end{gather}
Therefore, the average arrival rate of the two-state Markov fluid process is
\begin{gather}
r_{\avg} = \lambda P_{\on} = \lambda \, \frac{\alpha}{\alpha+\beta}. \label{eq:avgarrivaltwostate}
\end{gather}

\subsubsection{Markov Modulated Poisson Sources} \label{subsubsec:MMPP}

In this subsection, we assume that the data arrival to the buffer is a Poisson process whose intensity is controlled by a continuous-time Markov chain. For instance, the intensity of the Poisson arrival process is $\lambda_i$ in the $i^{\tth}$ state of the Markov chain. Therefore, the source arrival is modeled as a Markov-modulated Poisson process (MMPP). Assuming that the $\G$ is the irreducible transition rate matrix of the Markov chain and $\bLambda=diag\left\{\lambda_1, \lambda_2,\ldots, \lambda_n\right\}$ is the diagonal matrix of the intensities of the Poisson arrivals in different states, the effective bandwidth is given by \cite{elwalid}, \cite{ebw}
\begin{align}
a^*(\theta)=\frac{1}{\theta}\mu\left(\left(e^\theta-1\right)\bLambda+\G\right). \label{eq:poisEBW}
\end{align}

As in previous sections, we consider a two-state (ON/OFF) model in which there are no arrivals in the OFF state (i.e., the intensity is $0$) and the intensity of the Poisson arrival process is $\lambda$ bits/block in the ON state. Assuming the same generator matrix $\G$ as in \eqref{eq:generating}, we can express the effective bandwidth as
\begin{align}
\hspace{-0.3cm}a^{*}(\theta)=&\frac{1}{2\theta}\left[\left(e^\theta-1\right) \lambda-(\alpha+\beta)\right] \nonumber
\\
&+\frac{1}{2\theta}\sqrt{\big[\left(e^\theta-1\right) \lambda -(\alpha+\beta)\big]^2+4\alpha\left(e^\theta-1\right) \lambda} . \label{eq:2MMPPEBW}
\end{align}
Note that the average arrival rate in bits/block is again given by
\begin{gather}
r_{\avg} = \lambda P_{\on} = \lambda \, \frac{\alpha}{\alpha+\beta}. \label{eq:poisavgarrivaltwostate}
\end{gather}
We further note that if the transition rate $\beta = 0$, then we have $P_{\on} = 1$. In this case, MMPP model specializes to a pure Poisson source with intensity $\lambda$ bits/block, and the effective bandwidth of this source is given by
\begin{align}
\hspace{-0.3cm}a^{*}(\theta)=&\frac{1}{\theta}\left(e^\theta-1\right) \lambda. \label{eq:2PoissonEBW}
\end{align}

\section{Throughput with Markovian Source Models} \label{sec:metrics}

In this section, we formulate the throughput of wireless fading channels when the data arrivals are random and statistical queueing constraints are imposed. More specifically, we consider Markovian arrival models introduced in Section \ref{subsec:effective-bandw-of-sources}, namely discrete-time Markov sources, Markov fluids and Markov-modulated Poisson arrivals. The states in these Markov processes are differentiated by the corresponding arrival rates in these states, e.g., the arrival rate in the $i^{\tth}$ state is $\lambda_i$. If the stationary distribution of the Markov process is denoted by $\bpi$, the average arrival rate in an $n$-state Markov source model simply becomes
\begin{gather}
r_{\avg} = \sum_{i=1}^n \pi_i\lambda_i \label{eq:avgarrival}
\end{gather}
which is equal to the average departure rate when the queue is in steady state \cite{ChangZajic}.

We seek to determine the throughput by identifying the maximum average arrival rate that can be supported by the fading channel described in Section \ref{subsec:channelmodel} while satisfying the statistical QoS limitations given in the form in (\ref{eq:theta}). As shown in \cite[Theorem 2.1]{ChangZajic}, (\ref{eq:theta}) is satisfied, i.e., buffer violation probability decays exponentially fast with rate controlled by the QoS exponent $\theta$, if the effective bandwidth of the arrival process is equal to the effective capacity of the service process, i.e.,
\begin{align}
a^*(\theta) = C_E(\tsnr, \theta). \label{eq:equalityforQoS}
\end{align}
Hence by solving (\ref{eq:equalityforQoS}), we can determine the maximum average arrival rate $r_{\avg}^*(\tsnr, \theta)$. 
By specifying the effective bandwidth of different source models and incorporating the effective capacity of time-varying wireless transmissions in (\ref{eq:effcap}), the maximum average arrival rate can be determined for general $n$-state Markovian source models. Indeed, several $n$-state source models are addressed in Section \ref{sec:EE}. However, in our analysis in this section, to illustrate the impact of the arrival and system parameters in a lucid setting, we concentrate on the two-state (ON-OFF) arrival models and provide closed-form expressions for the maximum average arrival rates in terms of the source parameters and the effective capacity of the wireless transmissions. We also identify the characteristics of the throughput in the low-$\theta$ and high-SNR regimes. We note that the analysis throughout this section is applicable to any arbitrary fading correlation within each fading block, with the exception of high-SNR characterizations which are obtained under the assumption of i.i.d. fading.


\subsection{Discrete-Time Markov Sources} \label{subsec:d2Markov}

In this section, we consider two-state (ON/OFF) discrete Markov sources described in Section \ref{subsubsec:dMarkov}, and initially characterize the maximum average arrival rate $r_\avg^*$ that can be supported by the fading channel while satisfying the statistical QoS limitations given in the form in (\ref{eq:theta}).

\begin{theo} \label{theo:ravgdisctheo}
For the two-state (ON/OFF) discrete Markov source, the maximum average arrival rate (in bits/block) as a function of the QoS exponent $\theta$, effective capacity of the fading channel $C_E(\tsnr, \theta)$, and the state transition probabilities is expressed as
\begin{align}
r_\avg^*(\tsnr, \theta)=\frac{P_{\on}}{\theta}\log_e\left(\frac{e^{2\theta C_E(\ssnr, \theta)} - p_{11}e^{\theta C_E(\ssnr, \theta)}} {1-p_{11}-p_{22}+p_{22}e^{\theta C_E(\ssnr, \theta)}}\right). \label{eq:2discreteravg}
\end{align}
\end{theo}
\emph{Proof:} See Appendix \ref{subsec:ravgdisc}.

\begin{figure}
\begin{center}
\includegraphics[width=0.45\textwidth]{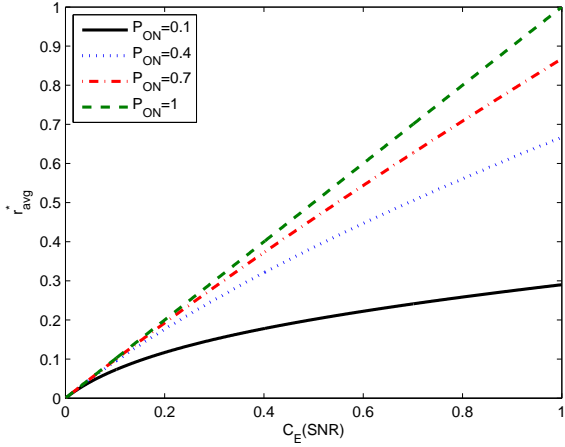}
\vspace{0.1cm}
\caption{Maximum average arrival rate $r^*_{\avg}$ vs. effective capacity $C_E(\tsnr)$ for different source statistics. No fading correlation, i.e., $\rho = 0$. $\theta = 1$.} \label{fig:ravgvsCE}
\end{center}
\end{figure}

Note that $r_\avg^*$ above is formulated in terms of the effective capacity, $C_E$, of wireless transmissions. In Fig. \ref{fig:ravgvsCE}, we plot the the maximum average arrival rate as a function of the effective capacity for different source characteristics when $\theta = 1$. It is easy to verify that when $P_{\on} = \frac{1-p_{11}}{2-p_{11} - p_{22}} = 1$ or equivalently $p_{22} = 1$, (\ref{eq:2discreteravg}) simplifies to $r_\avg^*(\tsnr, \theta) = C_E(\ssnr, \theta)$. Hence, when the source is always ON and therefore the arrivals are at a constant rate, maximum average arrival rate is equal to the effective capacity, as also observed in Fig. \ref{fig:ravgvsCE}. On the other hand, we notice in this figure that as $P_{\on}$ diminishes and the source becomes more bursty, throughput diminishes as well and smaller average arrival rates are supported for given effective capacity.

As also indicated in the above discussion and seen in (\ref{eq:2discreteravg}), $r_\avg^*(\tsnr, \theta)$ is in general a function of the state transition probabilities of the Markov arrival process in the presence of buffer constraints. On the other hand, as shown in the following result, this dependence disappears if no buffer constraints are imposed, i.e., when $\theta = 0$.

\begin{theo} \label{cor:discrete-Markov-theta-0}
As the statistical queueing constraints are relaxed by letting the QoS exponent $\theta$ approach zero, the maximum average arrival rate converges to
\begin{align}
\lim_{\theta \to 0} r_\avg^*(\tsnr, \theta) =& \sum_{i = 1}^m \E\left\{\log_2 (1 + \tsnr z_i)\right\} \text{ bits/block}. \label{eq:2discreteravg-theta-0}
\end{align}
Moreover, the first derivative of $r_\avg^*$ with respect to $\theta$ at $\theta = 0$ is
\begin{align}
\hspace{-0.1cm}\left.\frac{\partial r_\avg^*(\tsnr, \theta)}{\partial\theta}\right|_{\theta=0} \!\!=& -\frac{1}{2}\!\!\sum_{i,j=1}^m\!\cov\!\left\{\log_2 (1 \!+ \tsnr z_i),\log_2 (1 \!+ \tsnr z_j)\right\} \nonumber
\\
&-\frac{\eta }{2} \left(\sum_{i = 1}^m\E\left\{ \log_2 (1 + \tsnr z_i)\right\}\right)^2 \label{eq:dotravgdisctheta0}
\end{align}
where we define $\eta$ as
\begin{equation}
\eta = \frac{(1-p_{22})(p_{11}+p_{22})}{(1-p_{11})(2-p_{11}-p_{22})}. \label{eq:eta}
\end{equation}
\end{theo}

\emph{Proof:} See Appendix \ref{subsec:discrete-Markov-theta-0}.

We see from (\ref{eq:2discreteravg-theta-0}) that if no statistical buffer constraints are imposed i.e., if $\theta = 0$, then the maximum average arrival rate is equal to the ergodic capacity of the block-fading channel, and therefore is independent of the statistical characteristics of the discrete Markov arrival model. Moreover, the dependence of the maximum arrival rate in this regime on the channel statistics is only through the marginal distributions of the fading coefficients. Hence, channel correlation in each fading block does not play any role. However, this radically changes when $\theta >0$. For instance, we notice from (\ref{eq:dotravgdisctheta0}) that even with a small increase in $\theta$, $r_\avg^*$ starts varying with the source and channel statistics, as exemplified by the dependence of the first derivative on $\eta$ and the covariance function.

Having discussed the low-$\theta$ regime above, we next provide a characterization of $r_\avg^*(\tsnr, \theta)$ at high SNR values for i.i.d. Rayleigh fading.

\begin{theo} \label{cor:disc-high-SNR}
Assume that the channel fading coefficients are i.i.d. in each block and fading power $z = |h|^2$ is exponentially distributed with unit mean (i.e., Rayleigh fading is experienced). Then, we have
\begin{align}
\frac{1}{m} r_\avg^*(\tsnr, \theta) =
\left\{
\begin{array}{ll}
\frac{P_{\on}}{\theta \log_2e} \log_2 \tsnr + \mathcal{O}(1) & \text{if } \theta > \frac{1}{\log_2e}
\\
P_{\on}\log_2 \tsnr + \mathcal{O}(1) &\text{if } 0 < \theta < \frac{1}{\log_2e}
\\
\log_2 \tsnr + \mathcal{O}(1) & \text{if } \theta = 0
\end{array}\right. \label{eq:high-SNR-expansion-discretemarkov}
\end{align}
as $\tsnr \to \infty$.
\end{theo}
\emph{Proof:} See Appendix \ref{subsec:disc-high-SNR}.

Note that the high-SNR slope is defined as \cite{lozano}
\begin{align}
\mathcal{S}_{\infty} = \lim_{\ssnr \to \infty} \frac{\frac{1}{m} r_\avg^*(\ssnr, \theta)}{\log_2{\ssnr}}.
\end{align}
Theorem \ref{cor:disc-high-SNR} shows that the high-SNR slope of the maximum arrival rate for the two-state discrete Markov source that can be supported in the i.i.d Rayleigh fading channel is
\begin{align}
\mathcal{S}_{\infty} = \left\{
\begin{array}{ll}
\frac{P_{\on}}{\theta \log_2e} & \text{if } \theta > \frac{1}{\log_2e}
\\
P_{\on}&\text{if } 0 < \theta < \frac{1}{\log_2e}
\\
1 &\text{if } \theta = 0
\end{array}\right..
\end{align}
It is interesting to observe from Theorem \ref{cor:discrete-Markov-theta-0} that when no buffer constraints are imposed i.e., when $\theta = 0$, the high-SNR slope is $\mathcal{S}_{\infty} = 1$, again independent of the source statistics. On the other hand, when $\theta >0$, $\sinf$ becomes proportional to the ON probability and is now less than one unless the arrival rate is constant. Furthermore, for $\theta$ values greater than $\frac{1}{\log_2e}$, $\sinf$ starts decreasing with increasing $\theta$. Hence, the result in Theorem \ref{cor:disc-high-SNR} quantifies the performance degradation experienced at high SNR levels due to source randomness and statistical buffer constraints.

Let us further simplify the source model and set $p_{11}=1-s$ and $p_{22}=s$. The source is now described by the single parameter $s$. Notice that with this choice we have $P_{\on} = s$ and hence $s$ becomes a measure of the burstiness of the source. The smaller the $s$, the less frequently the data arrives and the more bursty the source becomes. At the other extreme, if $s =1$, source is ON all the time and we have constant arrival rate. Furthermore, with the above choice of $p_{11}$ and $p_{22}$, the expression for the maximum average arrival rate simplifies to
\begin{gather}
r_\avg^*(\tsnr, \theta)=\frac{s}{\theta}\log_e\left(\frac{e^{\theta C_E(\ssnr, \theta)} - (1-s)} {s}\right), \label{eq:ravgdiscreteq}
\end{gather}
which can readily be seen to be a diminishing function as $s$ decreases.
Therefore,  source burstiness generally hurts the throughput if we keep all other variables fixed.

\begin{figure}
\begin{center}
\includegraphics[width=0.45\textwidth]{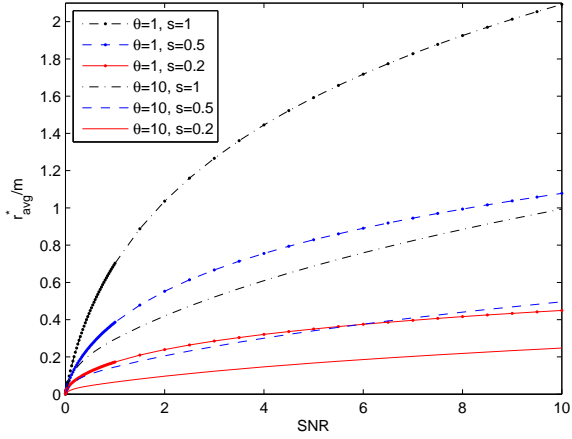}
\vspace{0.1cm}
\caption{Maximum average arrival rate $r^*_{\avg}$ vs. signal-to-noise ratio $\tsnr$ for different values of $\theta$ and source statistics. No fading correlation, i.e., $\rho = 0$.} \label{fig:ravgSNRdiscrete}
\end{center}
\end{figure}

We can further observe this in Fig. \ref{fig:ravgSNRdiscrete}, where we plot the maximum average arrival rate (or equivalently the throughput) as a function of SNR for different values of $s$ and the QoS exponent $\theta$. Numerical analysis verifies that as the source becomes more bursty with lower values of $s$, throughput diminishes. Conversely, throughput is maximized when $s = 1$ i.e., when we have constant arrival rates. It is also interesting to notice from (\ref{eq:ravgdiscreteq}) that the arrival rate in the ON state, which is given by $\lambda^*=\frac{r_\avg^*(\tsnr, \theta)}{s}$, increases as $s$ diminishes. Hence, smaller $s$ implies that data arrives less frequently but with bursts of increased rates. We also observe in Fig. \ref{fig:ravgSNRdiscrete} that the throughput reduction due to burstiness is more severe at high SNRs. This is indeed a consequence of the fact that high-SNR slope gets smaller as $P_{\on} = s$ decreases, as discussed above. Finally, we see in Fig. \ref{fig:ravgSNRdiscrete} that performance degradation is experienced as $\theta$ increases and hence stricter buffer constraints are imposed.

\begin{figure}
\begin{center}
\includegraphics[width=0.45\textwidth]{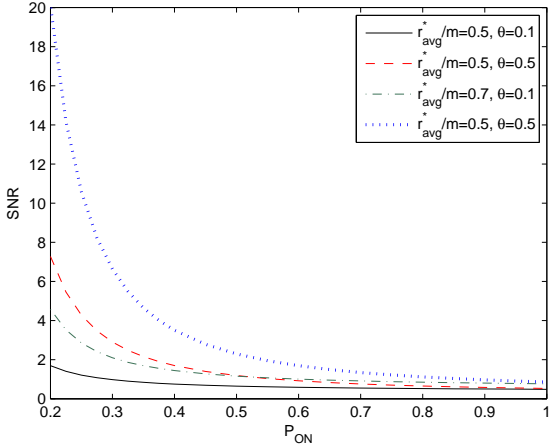}
\vspace{0.1cm}
\caption{Required SNR vs. ON probability, $P_{\on}$, for a given fixed average arrival rate. No fading correlation, i.e., $\rho = 0$.} \label{fig:SNRvsPONdiscrete}
\end{center}
\end{figure}

In Fig. \ref{fig:SNRvsPONdiscrete}, we plot the SNR levels required to support a given average arrival rate as a function of the ON-state probability for different values of the QoS exponent $\theta$. We observe that as $P_{\on}$ decreases and hence the source becomes more bursty, required SNR level increases in general. Interestingly, a sharper increase is experienced under stricter buffer constraints (e.g., when $\theta = 0.5$ rather than $\theta = 0.1$), indicating higher power/energy costs in these cases.

\begin{figure}
\begin{center}
\includegraphics[width=0.45\textwidth]{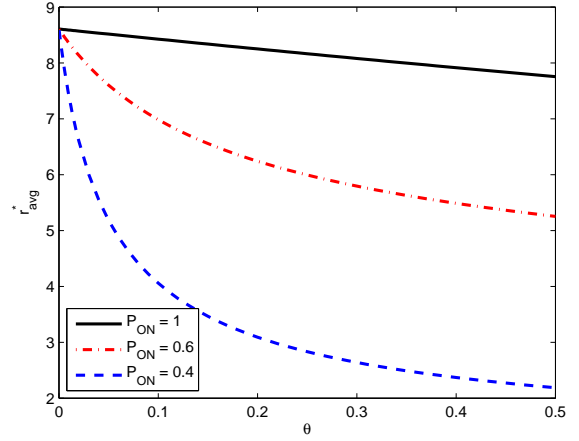}
\vspace{0.1cm}
\caption{Maximum average arrival rate $r^*_{\avg}$ vs. QoS exponent $\theta$ for different values of $P_{\on}$. No fading correlation, i.e., $\rho = 0$.} \label{fig:ravgvsthetadisc_burst}
\end{center}
\end{figure}

The low-$\theta$ regime is investigated in Fig. \ref{fig:ravgvsthetadisc_burst} where we plot the maximum average arrival rate $r^*_{\avg}$ vs. QoS exponent $\theta$ for different $P_{\on}$ values. We set $\tsnr=1$. We notice that all three curves converge to the same throughput value $r^*_{\avg}(0)$ as $\theta \to 0$, confirming the result in (\ref{eq:2discreteravg-theta-0}). Hence, source characteristics do not affect the throughput if no queuing constraints are imposed. As $\theta$ increases, throughput diminishes and the reduction in $r^*_{\avg}$ is more severe for more bursty sources (e.g., when $P_{\on} = 0.4$). We notice that, as predicted by (\ref{eq:dotravgdisctheta0}), this is already reflected by the different slopes of $r^*_{\avg}$ in the vicinity of $\theta = 0$. Hence, overall the system for more bursty sources becomes more cautious and supports smaller average arrival rates in order to avoid buffer overflows.

\begin{figure}
\begin{center}
\includegraphics[width=0.45\textwidth]{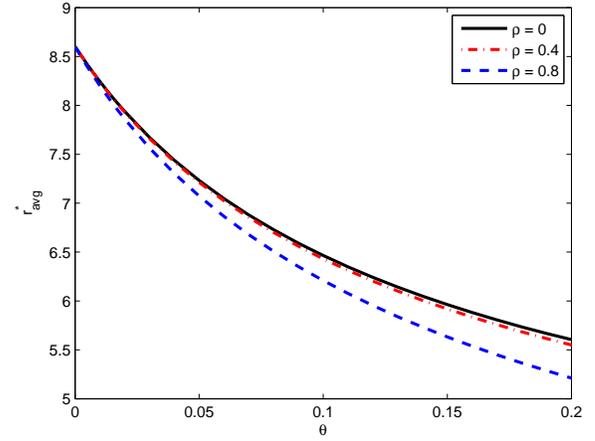}
\vspace{0.1cm}
\caption{Maximum average arrival rate $r^*_{\avg}$ vs. QoS exponent $\theta$ for different fading correlations. $P_{\on}=0.5$.} \label{fig:ravgvsthetadisc_cor}
\end{center}
\end{figure}

In Fig. \ref{fig:ravgvsthetadisc_cor}, we again plot the throughput as a function $\theta$ but for different values of $\rho$, which quantifies the correlation between fading coefficients in each fading block. We fix $\tsnr=1$ and set $P_{\on}=0.5$. Similar to burstiness, fading correlation does not have any effect on the throughput when $\theta=0$. When $\theta>0$, higher correlation (i.e., larger $\rho$) results in lower supported throughput under the same QoS constraints.

\begin{figure}
\begin{center}
\includegraphics[width=0.45\textwidth]{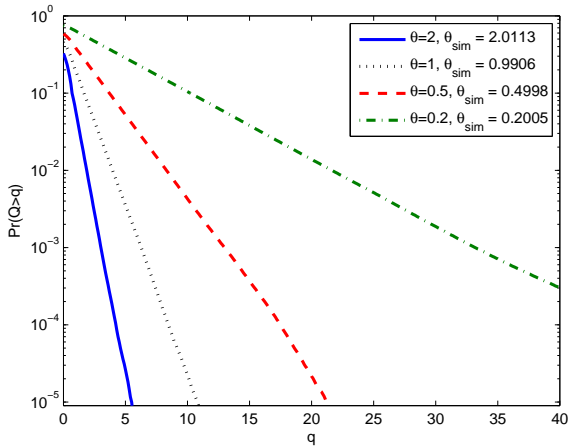}
\vspace{0.1cm}
\caption{Buffer overflow probability $\Pr\{Q>q\}$ vs. buffer threshold $q$ for different values of $\theta$. $p_{11} = p_{22} = 0.8$, SNR = $0$ dB} \label{fig:bufferqueue}
\end{center}
\end{figure}

\begin{figure}
\begin{center}
\includegraphics[width=0.45\textwidth]{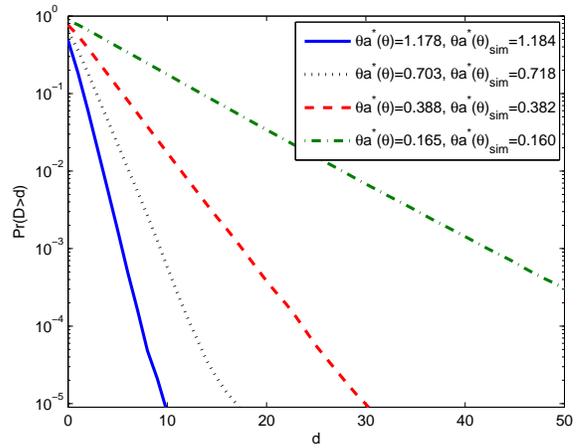}
\vspace{0.1cm}
\caption{Delay violation probability $\Pr\{D>d\}$ vs. delay threshold $d$ for different values of $\theta$. $p_{11} = p_{22} = 0.8$, SNR = $0$ dB} \label{fig:bufferdelay}
\end{center}
\end{figure}

Finally, we have conducted simulations to further verify the theoretical analysis and results. In particular, in the simulations, for fixed QoS exponent $\theta$, SNR, and state transition probabilities $p_{11}$ and $p_{22}$ of the ON/OFF discrete Markov source, we initially determine the maximum average arrival rate from (\ref{eq:2discreteravg}) and the corresponding maximum arrival rate in the ON state. Then, using the given statistical characterizations and the maximum arrival rate, we generate random Markov arrivals and assume that the arriving data is initially stored in the buffer before being transmitted. Transmission rates are simulated by generating realizations of i.i.d. Gaussian fading coefficients. Throughout this process, we track the queue evolution and the buffer state (i.e., the queue length) as the Markov arrivals occur (and hence more data gets stored) and transmissions at varying rates according to the generated fading coefficients are performed, clearing some data off the buffer. In Figs. \ref{fig:bufferqueue} and \ref{fig:bufferdelay}, we plot the simulated buffer overflow probability $\Pr\{Q \ge q\}$ and delay violation probability $\Pr\{D > d\}$, respectively, as functions of the corresponding thresholds, following $10^7$ runs of the simulation. We notice that while the theoretical analysis makes use of results from the theory of large deviations and is generally applicable for large thresholds, the simulation results are interestingly in excellent agreement with the theoretical predictions even at small values of the thresholds. For instance, we note from (\ref{eq:overflowprob-rev}) that $\log_e\Pr\{Q \ge q\} \approx \log_e \varsigma - \theta q$ and hence is expected to decay linearly in $q$ with slope $\theta$. We indeed observe this linear decay in Fig. \ref{fig:bufferqueue} (where the overflow probabilities are plotted in logarithmic scale) for even small to moderate values of $q$. Moreover, the slopes of the simulated curves, denoted by $\theta_{\text{sim}}$, are very close to the originally selected value of $\theta$. Similar conclusions apply to Fig. \ref{fig:bufferdelay} as well. In this figure, delay violation probabilities are determined by keeping track of the delay experienced by the data stored in the buffer until transmission. We again notice that the logarithm of the delay violation probability decays linearly with threshold $d$ (or equivalently the delay violation probability diminishes exponentially with $d$). Note that the slope of the linear decay is predicted from (\ref{eq:delayviolation}) to be $\theta a^*(\theta)$ where $a^*(\theta)$ is the effective bandwidth of the source. Again, the slope of the simulated curves are almost the same as this theoretical slope value, as indicated in the legend on the figure.

\subsection{Markov Fluid Sources} \label{subsec:f2Markov}

In this section, we consider Markov fluid sources. In the following, we go through similar steps as in the previous subsection and initially determine the maximum average arrival rates of ON/OFF Markov fluid sources that can be supported by the wireless channel as a function of the source transition rates and the effective capacity of wireless transmissions. Subsequently, we give characterizations of the maximum average arrival rates in the low-$\theta$ and high-SNR regimes.
\begin{theo} \label{theo:ravgfluidtheo}
For the two-state (ON/OFF) Markov fluid source, the maximum average arrival rate is given as
\begin{gather}
r_\avg^*(\tsnr, \theta)=P_{\on} \frac{\theta C_E(\ssnr, \theta) +\alpha+\beta} {\theta C_E(\ssnr, \theta) +\alpha} \, C_E(\tsnr, \theta). \label{eq:2fluidravg}
\end{gather}
\end{theo}
\emph{Proof:} See Appendix \ref{subsec:ravgfluid}.

Note that maximum average arrival rate generally depends on the transition rate matrix of the Markov fluid source. At the same time, similar to the discrete case, when there are no QoS constraints, source characteristics do not have any impact on the throughput. However, this changes drastically when $\theta > 0$ even if $\theta$ is vanishingly small. These properties are demonstrated analytically in the result below.
\begin{theo} \label{cor:fluid-Markov-theta-0}
As the statistical queueing constraints are relaxed by letting the QoS exponent $\theta$ approach zero, we have
\begin{align}
\lim_{\theta \to 0} r_\avg^*(\tsnr, \theta) =& \sum_{i = 1}^m \E\left\{\log_2 (1 + \tsnr z_i)\right\} \text{ bits/block}, \label{eq:2fluidravg-theta-0}
\intertext{and}
\hspace{-0.35cm}\left.\frac{\partial r_\avg^*(\tsnr, \theta)}{\partial\theta}\right|_{\theta=0} \!\!=& -\frac{1}{2}\!\!\sum_{i,j=1}^m\!\cov\!\left\{\log_2 (1 \!+ \tsnr z_i),\log_2 (1 \!+ \tsnr z_j)\right\} \nonumber
\\
&-\frac{\zeta}{2} \left(\sum_{i = 1}^m\E\left\{ \log_2 (1 + \tsnr z_i)\right\}\right)^2 \label{eq:dotravgfluidtheta0}
\end{align}
where $\zeta$ is defined as
\begin{equation}
\zeta=\frac{2\beta }{\alpha(\alpha+\beta)}. \label{eq:zeta}
\end{equation}
\end{theo}
\emph{Proof:} See Appendix \ref{subsec:fluid-Markov-theta-0}.

We note that when $\theta>0$, $r_\avg^*$ depends on the source and channel statistics. In \eqref{eq:dotravgfluidtheta0}, we observe the dependence of even the first derivative on channel correlations and source statistics via the covariance function and the parameter $\zeta$, respectively.

Next we present a high-SNR characterization of the throughput for Rayleigh fading.
\begin{theo} \label{cor:fluid-high-SNR}
Assume that the channel fading coefficients are i.i.d. in each block and fading power $z = |h|^2$ is exponentially distributed with unit mean (i.e., Rayleigh fading is experienced). Then, we have
\begin{align}
\frac{1}{m} r_\avg^*(\tsnr, \theta) =
\left\{
\begin{array}{ll}
\frac{P_{\on}}{\theta \log_2e} \log_2 \tsnr + \mathcal{O}(1) & \text{if } \theta > \frac{1}{\log_2e}
\\
P_{\on}\log_2 \tsnr + \mathcal{O}(1) &\text{if } 0 < \theta < \frac{1}{\log_2e}
\\
\log_2 \tsnr + \mathcal{O}(1) & \text{if } \theta = 0
\end{array}\right. \label{eq:high-SNR-expansion-fluidmarkov}
\end{align}
as $\tsnr \to \infty$.
\end{theo}
The proof of Theorem \ref{cor:fluid-high-SNR} is omitted due to its similarity to the proof of Theorem \ref{cor:disc-high-SNR} in Appendix \ref{subsec:disc-high-SNR}. Similar conclusions as in Section \ref{subsubsec:dMarkov} immediately apply.


Note that the throughput expression in \eqref{eq:2fluidravg} suggests that for sufficiently high SNR levels leading to $\theta C_E(\ssnr, \theta) \gg \alpha+\beta$, we have
\begin{gather}
r_\avg^*(\tsnr, \theta) \approx P_{\on}  C_E(\tsnr, \theta).
\end{gather}
Hence, at high SNRs, the maximum average arrival rate depends on the source statistics only through the ON probability. This is noted in the high-SNR behavior in (\ref{eq:high-SNR-expansion-fluidmarkov}) as well.

\begin{figure}
\begin{center}
\includegraphics[width=0.45\textwidth]{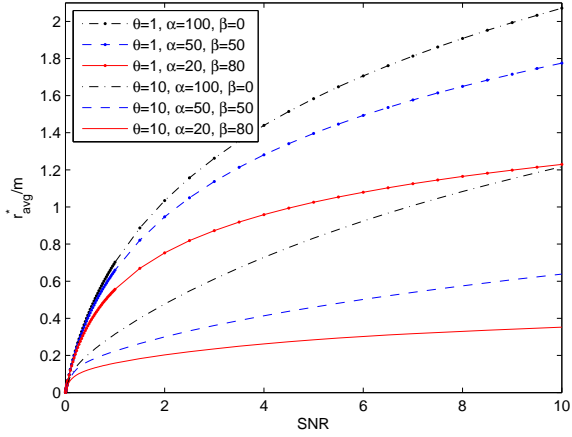}
\vspace{0.1cm}
\caption{Maximum average arrival rate $r^*_{\avg}$ vs. signal-to-noise ratio $\tsnr$ for different values of $\theta$ and source statistics. No fading correlation, i.e., $\rho = 0$.} \label{fig:ravgSNRfluid}
\end{center}
\end{figure}
\begin{figure}
\begin{center}
\includegraphics[width=0.45\textwidth]{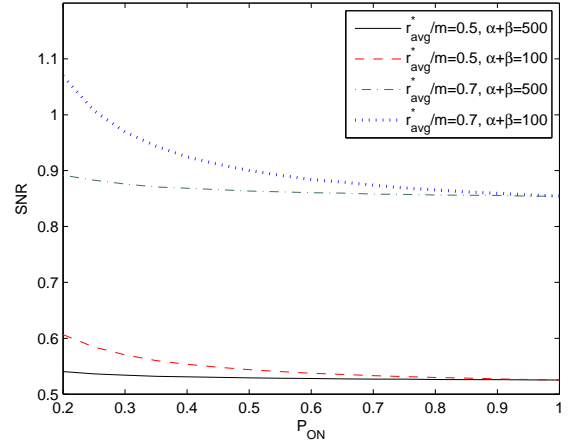}
\vspace{0.1cm}
\caption{Required SNR vs. ON probability, $P_{\on}$,  for a given average arrival rate. $\theta=0.5$. No fading correlation, i.e., $\rho = 0$.} \label{fig:SNRvsPONfluid}
\end{center}
\end{figure}

In Fig. \ref{fig:ravgSNRfluid}, we plot $r_\avg^*$ vs. SNR curves for different $\alpha$, $\beta$, and $\theta$ values. We immediately observe that throughput diminishes with increasing $\theta$ and decreasing $P_{\on}$.
In Fig. \ref{fig:SNRvsPONfluid}, we analyze the effect of $r_{\avg}$, $P_{\on}$, and $\alpha+\beta$ on the required SNR levels. For Markov fluid sources, ON state probability is not the sole indicator of burstiness. Having low $\alpha$ and $\beta$ values also indicates that source is more bursty as the transition between ON and OFF states becomes less frequent. Hence, OFF state can be more persistent. When $\alpha$ and $\beta$ are large, state transitions occur more rapidly, leading to lower required SNR levels. Again, we notice that the burstiness is harmful for the system.

\begin{figure}
\begin{center}
\includegraphics[width=0.45\textwidth]{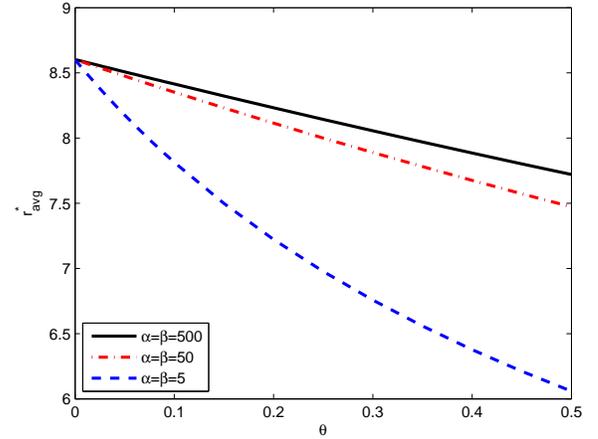}
\vspace{0.1cm}
\caption{Maximum average arrival rate $r^*_{\avg}$ vs. QoS exponent $\theta$ for different values of $\alpha$ and $\beta$. $P_{\on} = 0.5$, $\rho = 0$, and $\tsnr = 0$ dB.} \label{fig:ravgvsthetafluid_burst}
\end{center}
\end{figure}

In Fig. \ref{fig:ravgvsthetafluid_burst}, we plot the maximum average arrival rate $r^*_{\avg}$ as a function of $\theta$ for different values of $\alpha$ and $\beta$. Notice that by keeping $\alpha = \beta$, the ON probability $P_{\on}$ is fixed at 0.5, while average durations of ON and OFF states vary as the values of $\alpha = \beta$ change. For example, higher $\alpha$ and $\beta$ values lead to shorter periods for ON and OFF states on average. As an outcome of this fact, we observe in the figure that higher throughput is achieved with sources having higher $\alpha+\beta$. 

\subsection{Markov Modulated Poisson Sources} \label{subsec:MMPP2}


In this section, we address two-state (ON/OFF) MMPP sources. Similarly as for the previous source models, we determine the maximum average arrival rate of the MMPP source, which can be supported by the fading channel in the presence of QoS constraints, and investigate the throughput in the low-$\theta$ and high-SNR regimes. The results can be immediately specialized to pure Poisson sources by setting $\beta = 0$.

\begin{theo} \label{theo:ravgMMPPtheo}
For the two-state (ON/OFF) MMPP source model, the maximum average arrival rate is
\begin{gather}
r_\avg^*(\tsnr, \theta)=P_{\on} \frac{\theta\left[\theta C_E(\ssnr, \theta) +\alpha+\beta\right]} {(e^\theta\!\!-\!1)\left[\theta C_E(\ssnr, \theta) +\alpha\right]} \, C_E(\tsnr, \theta). \label{eq:2MMPPravg}
\end{gather}
\end{theo}
\emph{Proof:} See Appendix \ref{subsec:ravgMMPP}.

It is interesting to observe that the throughput with the MMPP source is almost identical to that with the Markov fluid source model, save only for the multiplicative factor $\frac{\theta}{e^{\theta} -1}$ in (\ref{eq:2MMPPravg}). Note that $\frac{\theta}{e^{\theta} -1} < 1$ for $\theta >0$ and diminishes exponentially fast with increasing $\theta$. Hence, the throughput is generally smaller with MMPP sources and decreases fast with $\theta$. This can be attributed to the much more randomness/burstiness we experience with an MMPP source with respect to the previous Markov models. Note that the arrival rate in the ON state, rather than being a constant as in the previous cases, is determined by a Poisson process. Hence, the presence of the term $\frac{\theta}{e^{\theta} -1}$ is due to this Poisson property. Indeed, if we have a pure Poisson source, the maximum average arrival rate is  $r_\avg^*(\tsnr, \theta)=\frac{\theta} {(e^\theta-1)} \, C_E(\tsnr, \theta)$ obtained by setting $\beta = 0$. The cost of this additional randomness is reflected in the following results as well.

\begin{theo} \label{cor:MMPP-Markov-theta-0}
As the statistical queueing constraints are relaxed by letting the QoS exponent $\theta$ approach zero, we have
\begin{align}
\lim_{\theta \to 0} r_\avg^*(\tsnr, \theta) =& \sum_{i = 1}^m \E\left\{\log_2 (1 + \tsnr z_i)\right\} \text{ bits/block}, \label{eq:2MMPPravg-theta-0}
\\ \intertext{and}
\hspace{-0.35cm}\left.\frac{\partial r_\avg^*(\tsnr, \theta)}{\partial\theta}\right|_{\theta=0} \!\!=& -\frac{1}{2}\!\!\sum_{i,j=1}^m\!\cov\!\left\{\log_2 (1 \!+ \tsnr z_i),\log_2 (1 \!+ \tsnr z_j)\right\} \nonumber
\\
&-\frac{\zeta}{2} \left(\sum_{i = 1}^m\E\left\{ \log_2 (1 + \tsnr z_i)\right\}\right)^2 \nonumber
\\
&-\frac{1}{2}\sum_{i = 1}^m \E\left\{\log_2 (1 + \tsnr z_i)\right\}\label{eq:dotMMPPfluidtheta0}
\end{align}
where
\begin{equation}
\zeta=\frac{2\beta }{\alpha(\alpha+\beta)}.
\end{equation}
\end{theo}

\emph{Proof:} See Appendix \ref{subsec:MMPP-Markov-theta-0}.

When the system is free of QoS limitations, the maximum average arrival rate for the MMPP source again turns out to be equal to the ergodic capacity. However, the throughput has a steeper decline in the low-$\theta$ regime due to the third term on the right-hand side of (\ref{eq:dotMMPPfluidtheta0}).

\begin{theo} \label{cor:mmpp-high-SNR}
Assume that the channel fading coefficients are i.i.d. in each block and fading power $z = |h|^2$ is exponentially distributed with unit mean (i.e., Rayleigh fading is experienced). Then, we have
\begin{align}
\frac{1}{m} r_\avg^*(\tsnr, \theta)\! =\!
\left\{\!\!\!\!
\begin{array}{ll}
\frac{P_{\on}}{\left(e^\theta-1\right) \log_2\!e} \log_2 \!\tsnr \!+\! \mathcal{O}(1)\! & \text{if } \theta \!>\! \frac{1}{\log_2e}
\\
\frac{\theta}{e^\theta-1 }P_{\on}\log_2 \tsnr + \mathcal{O}(1) &\text{if } 0 \!< \theta \!<\! \frac{1}{\log_2e}
\\
\log_2 \tsnr + \mathcal{O}(1) & \text{if } \theta\! = 0
\end{array}\right. \label{eq:high-SNR-expansion-MMPP}
\end{align}
as $\tsnr \to \infty$.
\end{theo}

Since the ratio between the MMPP throughput and Markov fluid throughput always stays at $\frac{\theta}{e^\theta-1 }$, we can immediately obtain the above high-SNR characterization, using the formulations in \eqref{eq:high-SNR-expansion-fluidmarkov}.

\begin{figure}
\begin{center}
\includegraphics[width=0.45\textwidth]{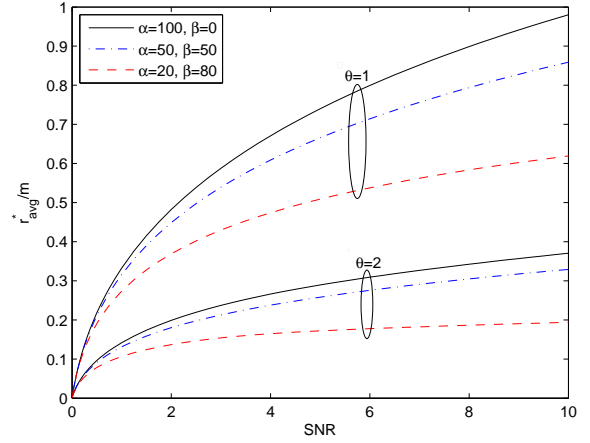}
\vspace{0.1cm}
\caption{Maximum average arrival rate $r^*_{\avg}$ vs. signal-to-noise ratio $\tsnr$ for different values of $\theta$ and different source statistics. No fading correlation, i.e., $\rho = 0$.} \label{fig:ravgSNRMMPP}
\end{center}
\end{figure}

\begin{figure}
\begin{center}
\includegraphics[width=0.45\textwidth]{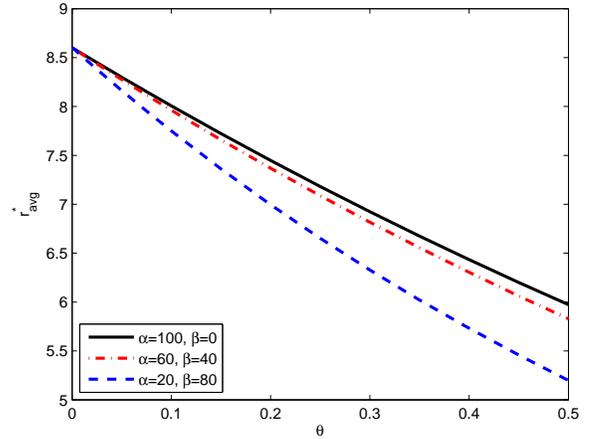}
\vspace{0.1cm}
\caption{Maximum average arrival rate $r^*_{\avg}$ vs. QoS exponent $\theta$ for different source statistics. $\rho = 0$.} \label{fig:ravgvsthetaMMPP_burst}
\end{center}
\end{figure}

In the numerical results, we have similar conclusions as in the Markov fluid case. The primary difference is the reduced throughput for given $\theta$, which, for instance, is readily seen when we compare Figs. \ref{fig:ravgSNRfluid} and \ref{fig:ravgSNRMMPP}, where we have throughput vs. SNR curves for Markov fluid and MMPP sources, respectively.
In Fig. \ref{fig:ravgvsthetaMMPP_burst}, we display the maximum average arrival rate $r^*_{\avg}$ as a function of $\theta$. We set $\alpha+\beta=100$ and $\tsnr=1$, and vary $\alpha$ and $\beta$ and hence the ON probability. We note that as $P_{\on}$ decreases, the performance degrades faster with increasing $\theta$, as indicated by the steeper slopes.

\subsection{Comparative View of Source Models and Performance Levels}

In our analysis, we have considered discrete-time Markov, Markov fluid, and MMPP arrival models. All models possess the Markovian property in the sense that the evolution of the Markov chains and hence the state transitions satisfy the Markov condition and are described by the transition probability matrix in the case of discrete-time models and by the transition rate matrix in the case of fluid (or equivalently continuous-time) models. Also, state holding times are geometrically distributed in discrete-time models and exponentially distributed in continuous-time models, and hence exhibit the memoryless property.

At the same time, there are distinct differences between different source models. For instance, transitions between states occur in discrete time steps in discrete-time Markov models while the Markov chain can spend a continuous amount of time in any state in Markov fluid models (i.e., the length of time spent in any state is a continuous random variable or more explicitly holding times are exponentially distributed as also noted above). MMPP models are further differentiated. In the discrete-time Markov and Markov fluid models, arrival rates are assumed to be constant in any given state. On the other hand, when the arrivals are modeled as MMPP, arrival rate is Poisson distributed in each state with a different intensity. Hence, MMPP sources exhibit a higher level of variation in this sense and can be regarded as a more bursty source.

We also remark that ON/OFF discrete-time Markov and Markov fluid source models can be easily specialized to the source with a constant arrival rate by letting ON state probability $P_\on = 1$. On the other hand, when $P_\on = 1$ in the ON/OFF MMPP source, we have a pure Poisson arrival source.

Finally, we note that although there is a certain degree of similarity in the analysis of discrete-time Markov and Markov fluid models and their throughput performances (e.g., high-SNR characterizations are the same in Theorems \ref{cor:disc-high-SNR} and \ref{cor:fluid-high-SNR}), the set of results for one model do not immediately follow from those for the other model as seen in the throughput formulations in (\ref{eq:2discreteravg}) and (\ref{eq:2fluidravg}) and the definitions of $\eta$ and $\zeta$ in (\ref{eq:eta}) and (\ref{eq:zeta}), respectively. However, there is a clear distinction when MMPP sources are considered. As also discussed in Section \ref{subsec:MMPP2}, higher level of burstiness of MMPP sources penalizes the performance, and lower throughput levels are achieved in general with these sources.


\section{Energy Efficiency Analysis} \label{sec:EE}

In this section, we conduct a low-SNR analysis and investigate the energy efficiency in fading channels when data arrivals are random and statistical queueing constraints are imposed.
We first identify the energy efficiency metrics. Subsequently, we consider different source arrival models and provide closed-form expressions for the energy efficiency metrics when the arrival rate is constant or follows a two-state Markovian model. We also numerically analyze specific $n$-state Markovian sources. Similarly as in the previous section, arbitrary fading correlation within each fading block is considered in the analysis. 


\subsection{Energy Efficiency Metrics} \label{subsec:EEmetrics}
Before defining the energy efficiency metrics, we briefly describe the concavity of the maximum average arrival rate as a function of SNR in the two-state (ON/OFF) arrival models (or if the arrival rates in an $n$-state model can be expressed as multiples of a certain single rate). In \cite[Lemma 1]{gursoy-Twireless09}, it was proven that effective capacity is a concave function of SNR. Elwalid and Mitra \cite{elwalid} showed that the effective bandwidth of a source is monotonically increasing when any arrival rate $\lambda_i$ increases and is convex in the arrival rates $\{\lambda_1, \lambda_2, \ldots, \lambda_N\}$. In the ON/OFF arrival models, we have a single arrival rate $\lambda$. Since effective bandwidth is a monotonically increasing and convex function of $\lambda$, the inverse function of the effective bandwidth ${a^*}^{-1}$ exists and is a nondecreasing concave function. More specifically, the maximum arrival rate can be expressed as $\lambda^*(\tsnr, \theta)={a^*}^{-1}\left(C_E(\tsnr, \theta)\right)$, which is a nondecreasing concave function of the effective capacity, which is concave in SNR. Using the composition properties of concave functions \cite{co-boyd}, we realize that the maximum arrival rate is concave in SNR. Thus, the maximum average arrival rate $r_\avg^*(\tsnr, \theta)$ is also concave in SNR.

In our analysis, following the approach in \cite{verdu}, we study the minimum energy per bit and the wideband slope, which is defined as the
slope of the spectral efficiency curve at zero spectral efficiency, as the performance metrics of energy efficiency.
While minimum bit energy is a performance measure in the limit as $\tsnr \to
0$ (due to the concavity of the throughput), wideband slope has emerged as a tool that enables us to analyze
the energy efficiency at low but nonzero $\tsnr$ levels.
In our setup, we define energy per bit as
\begin{gather}
\frac{E_b}{N_0} = \frac{\tsnr}{ r_\avg^*(\tsnr,\theta)/m }
\end{gather}
where the normalization with $m$ is due to our assumption that $r_\avg^*$ is in the units of bits per $m$ symbols (or equivalently per block).

The minimum energy per bit $\frac{E_b}{N_0}_{\tmin}$ under QoS
constraints can be obtained from
\begin{equation}\label{eq:ebnomin-ra}
\frac{E_b}{N_0}_{\tmin}=\lim_{\tsnr\rightarrow0}\frac{\tsnr}{r_\avg^*(\tsnr,\theta)/m}=\frac{1}{ \dot{r}_\avg^*(0)/m }.
\end{equation}
At $\frac{E_b}{N_0}_{\tmin}$, the slope $\mathcal {S}_0$ of the
throughput versus $E_b/N_0$ (in dB) curve is defined as
\cite{verdu}
\begin{equation}\label{slope}
\mathcal{S}_0=\lim_{\frac{E_b}{N_0}\downarrow\frac{E_b}{N_0}_\tmin}
\frac{r_\avg^*(\tsnr,\theta)/m}{10\log_{10}\frac{E_b}{N_0}-10\log_{10}\frac{E_b}{N_0}_\tmin}10\log_{10}2.
\end{equation}
The
wideband slope can also be found from
\begin{equation}\label{eq:widebandslope-ra}
\mathcal{S}_0=-\frac{2\big(\dot{r}_\avg^*(0)/m\big)^2} { \ddot{r}_\avg^*(0)/m}\log_e{2}
\end{equation}
where $\dot{r}_\avg^*(0)$ and $\ddot{r}_\avg^*(0)$ are the first and second
derivatives, respectively, of the function $r_\avg^*(0)$ with respect to SNR at zero SNR. $\frac{E_b}{N_0}_{\tmin}$
and $\mathcal{S}_0$ essentially provide a linear approximation of the throughput curve at low SNR levels.

\subsection{Energy Efficiency with Constant Arrival Rate} \label{subsec:constantrate}

In this section, we assume that the source arrival rate is fixed. Hence, we investigate the energy efficiency in the absence of source randomness and examine the impact of fading correlation and queueing constraints. As discussed in the previous section, effective capacity, $C_E(\tsnr,\theta)$, characterizes the maximum constant arrival rate in the presence of QoS constraints described by the QoS exponent $\theta$. Hence, we in this case have $r_\avg^*(\tsnr, \theta) = C_E(\tsnr, \theta)$. In the following result, we provide the minimum bit energy and wideband slope expressions under these assumptions.

\begin{theo} \label{theo:correlatedfading}
Assume that the source arrival rate is constant.
Then, the minimum energy per bit and wideband slope expressions as a function of the QoS exponent $\theta$ are given, respectively, by
\begin{gather}
{\frac{E_b}{N_0}}_{\text{min}} = \frac{\log_e2}{\E\left\{ z\right\}} \label{eq:ebno-min}
\intertext{and}
\so = \frac{2(\E\left\{ z \right\})^2} {\frac{\theta}{m\log_e2}\sum_{i,j=1}^{m}\cov\left\{ z_i, z_j\right\} + \E\left\{ z^2\right\} } \label{eq:widebandslope}
\end{gather}
where $\cov(z_i,z_j) = \E\{z_iz_j\} - \E\{z_i\}\E\{z_j\}$ is the covariance of $z_i$ and $z_j$.
\end{theo}

\emph{Proof:} See Appendix \ref{subsec:constantarrival}.


%
%
%

\begin{rem}
As can be seen in (\ref{eq:ebno-min}), the minimum energy per bit, which is achieved in the asymptotic regime in which SNR vanishes, does not depend on the QoS exponent $\theta$, hence is not affected by the presence of the buffer limitations. Indeed, this is the fundamental limit in Gaussian channels \cite{verdu}. Wideband slope $\so$, on the other hand, depends on the QoS constraints via the QoS exponent $\theta$. It can be easily seen that higher the value of $\theta$, the stricter the QoS constraints are and the smaller the value of the wideband slope is, indicating the increased energy requirements. Furthermore, it can be readily verified that wideband slope decreases with increased fading correlation. Or conversely, variations in the channel conditions are favorable for improved energy efficiency.
\end{rem}

\begin{figure}
\begin{center}
\includegraphics[width=0.45\textwidth]{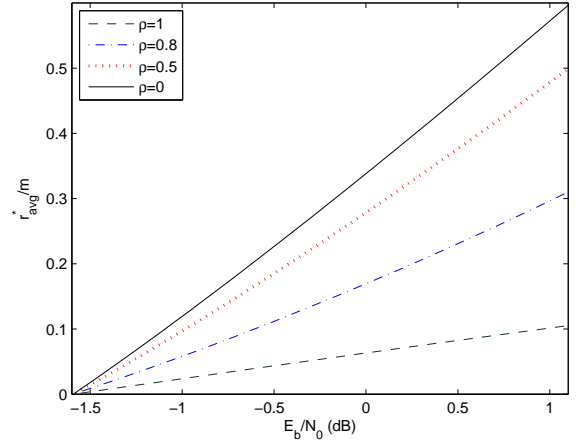}
\caption{Normalized maximum average arrival rate $\frac{1}{m} r_{\avg}^*$ vs. energy per bit $\frac{E_b}{N_0}$ for different fading correlations. $\theta = 1$, $m=10$.}\label{fig:figcorrelation}
\end{center}
\end{figure}

In Fig. \ref{fig:figcorrelation}, we plot the normalized maximum average arrival rate $\frac{1}{m} r_{\avg}^*$ as a function of the energy per bit $\frac{E_b}{N_0}$ for different correlation factors $\rho$ when $\theta = 1$ and $\E\{z\} = \E\{|h|^2\} = \sigma_h^2 = 1$. As predicted by Theorem \ref{theo:correlatedfading}, all curves converge to the same minimum energy per bit of $\frac{E_b}{N_0}_{\text{min}} = \frac{\log_e2}{\E\left\{ z\right\}} = \log_e2 = -1.59$ dB as $\tsnr$  and hence $r_{\avg}^*$ vanish. On the other hand, wideband slopes are different for different values of $\rho$. As discussed above, as $\rho$ and hence correlation diminishes from 1 to 0, slopes increase progressively. It is also interesting to note that in the absence of QoS constraints, i.e., when $\theta = 0$, such a distinction disappears. The wideband slope becomes $\so = \frac{2(\E\left\{ z \right\})^2} {\E\left\{ z^2\right\}}$, which clearly does not depend on the fading correlation.

\subsection{Energy Efficiency with Discrete-Time Markov Sources} \label{subsec:discrete}

Starting with this subsection, we incorporate random arrivals into our energy efficiency analysis and determine how source randomness affects the performance.

\subsubsection{ON-OFF Discrete-Time Markov Sources}
We assume that data arrival is either in the ON or OFF state in each block duration of $m$ symbols. As we have previously stated, in the ON state, $\lambda$ bits arrive (i.e., the arrival rate is $\lambda$ bits/block) while there are no arrivals in the OFF state. Below, we provide our results on energy efficiency.

\begin{theo} \label{theo:discrete}
Assume that the source arrival rate is random and follows the described discrete-time ON-OFF Markov model.
Then, the minimum energy per bit and wideband slope expressions as a function of the QoS exponent $\theta$ are given, respectively, by
\begin{gather}
{\frac{E_b}{N_0}}_{\text{min}} = \frac{\log_e2}{\E\left\{ z\right\}} \label{eq:ebnomin-ra-theo}
\intertext{and}
\hspace{-.4cm}\mathcal{S}_0=\frac{2(\E\left\{ z \right\})^2} {  \eta\frac{\theta m}{\log_e2}(\E\left\{ z \right\})^2 +\frac{\theta}{m\log_e2}\sum_{i,j=1}^m \cov\left\{ z_i, z_j\right\} + \E\left\{ z^2 \right\} } \label{eq:widebandslope-ra-theo}
\end{gather}
where $\eta$ is defined in \eqref{eq:eta}.
\end{theo}

\emph{Proof:} See Appendix \ref{subsec:discretearrival}.

Interestingly, ${\frac{E_b}{N_0}}_{\text{min}}$ again turns out to be a very robust quantity. Regardless of the buffer constraints, channel correlations, and randomness of the arrivals, the minimum received energy per bit is ${\frac{E_b}{N_0}}_{\text{min}} = \log_e2 = -1.59$ dB when $\E\{z\} =1$. On the other hand, the impact of random arrivals on the wideband slope is perspicuous in \eqref{eq:widebandslope-ra-theo}. When compared to \eqref{eq:widebandslope}, we immediately notice that having random arrivals leads to the introduction of the term $\eta\frac{\theta m}{\log_e2}(\E\left\{ z \right\})^2$ in the denominator of (\ref{eq:widebandslope-ra-theo}). Notice that when $p_{22} = 1$ and $p_{11} = 0$, we have $P_{\on} = 1$, meaning that we have a constant arrival rate. In this case, $\eta = 0$ and indeed (\ref{eq:widebandslope-ra-theo}) specializes to (\ref{eq:widebandslope}). More generally, we have $\eta \ge 0$ for all $p_{11}, p_{22}  \in [0,1]$. Therefore, random arrivals potentially decreases the wideband slope and increases the energy requirements.

\begin{figure}
\begin{center}
\includegraphics[width=0.45\textwidth]{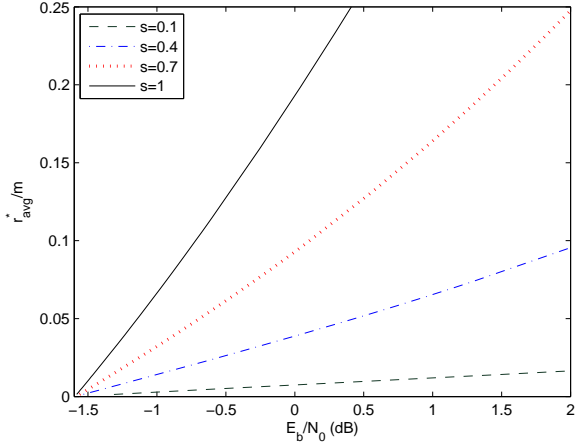}
\caption{Maximum average arrival rate $\frac{1}{m} r_{\avg}^*$ vs. energy per bit $\frac{E_b}{N_0}$ for different values of $P_{\on}=s$ when $\rho=0.75$, $\theta = 1$, and $m=10$.}\label{fig:figsourcevariation}
\end{center}
\end{figure}

This is more clearly seen again in the special case in which $p_{11}=1-s$ and $p_{22}=s$. Now, we have $\eta=\frac{p_{11}}{p_{22}}=\frac{1-s}{s}$ and the wideband slope is
\begin{align}
\mathcal{S}_0=\frac{2(\E\left\{ z \right\})^2} {  \frac{1-s}{s}\frac{\theta m}{\log_e2}(\E\left\{ z \right\})^2 +\frac{\theta}{m\log_e2}\sum_{i,j=1}^m \cov\left\{ z_i, z_j\right\} + \E\left\{ z^2 \right\} }. \label{eq:on-offchannel}
\end{align}
As $P_{\on} = s$ decreases, the wideband slope decreases as well. Therefore, the source becoming more bursty leads to increased energy per bit. This is illustrated in Fig. \ref{fig:figsourcevariation} where maximum average arrival rate vs. energy per bit is plotted and the same channel fading and correlation model as in Fig. \ref{fig:figcorrelation} is used. In this figure, we assume $\theta = 1$ and $\rho = 0.75$. As predicted, the minimum bit energies are all the same. However, we have diminishing slopes with decreasing $P_{\on}$. Note that for a fixed average arrival rate, as $P_{\on}$ gets smaller, source becomes more bursty. Data arrives less frequently but with a higher rate. This in turn increases energy per bit as implied by smaller wideband slopes.


\subsubsection{Discrete-Time Markov Sources with $n$ States}
\begin{figure}
\begin{center}
\includegraphics[width=0.45\textwidth]{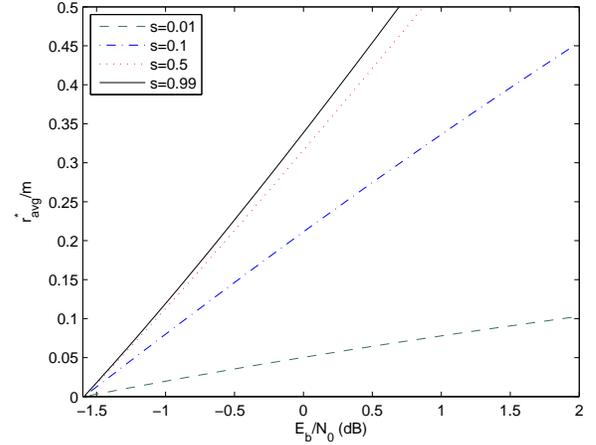}
\caption{Maximum average arrival rate $\frac{1}{m}r_{\avg}^*$ vs. energy per bit $\frac{E_b}{N_0}$ for different values of $s$ when channel blocks are uncorrelated and $\theta = 1$. Number of states for the arrival process is $n = 10$.}\label{fig:nstate_q}
\end{center}
\end{figure}

In this model, we assume that there are $n-1$ sources, each having its own ON and OFF states. In the ON state, a source
sends data to the buffer at the rate of $\lambda$ bits/block. Otherwise, it is in OFF state in which no data is generated. In this set-up, depending on how many sources are active (i.e., are in ON state), data arrivals to the buffer can be regarded as a discrete-time Markov process with $n$ states. In the $i^{th}$ state of this model, $(i-1)$ sources are active. For simplicity, we assume that the probability of each source being active in a given block is $s$, independent of the previous states and of the other sources. Then, the state probabilities will be given by
\begin{equation}
\pi_i=\binom{n-1}{i-1}s^{i-1}(1-s)^{n-i} \text{ for } i=1, 2, \ldots, n. \label{eq:binomialpdf}
\end{equation}
Note that the system is essentially memoryless because each state is independent of the previous state. Hence, transition probability matrix becomes
\begin{align}
\J=
\begin{bmatrix}
 \pi_1 & \pi_2 &\cdots&\pi_n\\
 \pi_1 & \pi_2 &\cdots&\pi_n \\
 \vdots &\vdots&\vdots& \vdots\\
 \pi_1 & \pi_2 &\cdots&\pi_n \label{eq:birthdeathG}
\end{bmatrix}.
\end{align}
Using \eqref{eq:binomialpdf} we can write the average rate expression as
\begin{equation}
r_{\avg}=\sum_{i=1}^n \binom{n-1}{i-1}q^{i-1}(1-q)^{n-i}(i-1)\lambda = (n-1)q\lambda. \label{eq:memorylessdiscreteavgrate}
\end{equation}
For this case, we do not have closed-form expressions. However, we can easily obtain the effective bandwidth and maximum average arrival rate numerically. In particular, by numerically solving \eqref{eq:equalityforQoS} and using \eqref{eq:memorylessdiscreteavgrate}, we can determine the maximum average arrival rate $r_{\avg}^*$ as a function of SNR. In Fig. \ref{fig:nstate_q}, we display the maximum average arrival rate as a function of energy per bit. Similarly as in the simple ON/OFF model, we observe that when source burstiness is decreased by increasing $s$, energy efficiency improves.

\begin{figure}
\begin{center}
\includegraphics[width=0.45\textwidth]{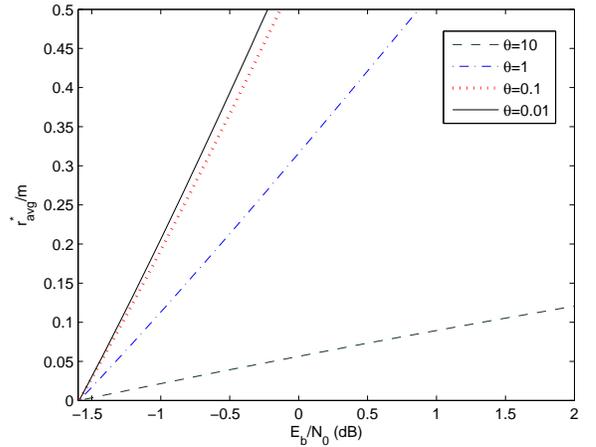}
\vspace{0.1cm}
\caption{Maximum average arrival rate $\frac{1}{m}r_{\avg}^*$ vs. energy per bit $\frac{E_b}{N_0}$ for different $\theta$ values when channel blocks are uncorrelated and $s = 0.5$. Number of states for arrival process is $n = 10$.}\label{fig:nstate_theta}
\end{center}
\end{figure}

To have a better understanding of the effect of the QoS constraints, average arrival rate curves in Fig. \ref{fig:nstate_theta} are obtained for different $\theta$ values. We first notice that QoS exponent $\theta$ does not have any effect on the minimum energy per bit because all curves merge at $-1.59$ dB which is again the minimum energy per bit for all $\theta$ values. However, energy efficiency degrades with stricter QoS conditions  as increasing $\theta$ reduces the wideband slope.

\begin{figure}
\begin{center}
\includegraphics[width=0.45\textwidth]{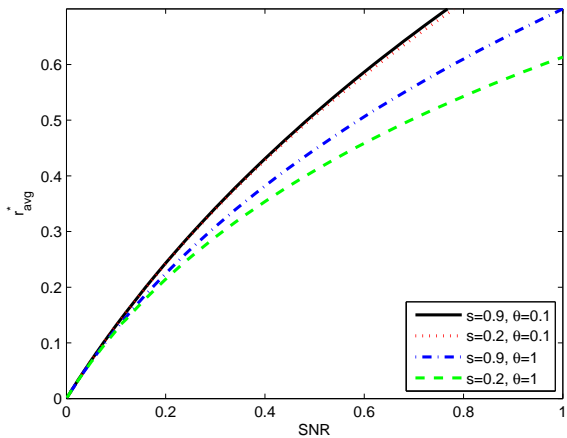}
\vspace{0.1cm}
\caption{Maximum average arrival rate $r_{\avg}^*$ vs. signal-to-noise ratio $\tsnr$ when channel blocks are uncorrelated. Number of states of the arrival process is $n = 10$.}\label{fig:nstate_dravgSNR}
\end{center}
\end{figure}

Finally, for comparison purposes, we depict the throughput as a function of SNR in Fig. \ref{fig:nstate_dravgSNR} for different source characteristics and $\theta$ values. The trends in the throughput vs. SNR curves for the considered $n$-state discrete Markov source are observed to be similar to those in Fig. \ref{fig:ravgSNRdiscrete} plotted for the ON-OFF discrete Markov source. For instance, again increased source burstiness (i.e., lower values of s) and stricter queueing constraints (i.e., higher $\theta$ values) result in the degradation of the throughput for both two-state (ON/OFF) and $n$-state source models.

\subsection{Energy Efficiency with Markov Fluid Sources} \label{subsec:fluid}

\subsubsection{ON-OFF Markov Fluid Model}
Now, we consider Markov fluid sources with two states, namely OFF state with no arrivals and ON state in which the arrival rate is $\lambda$. The generating matrix is defined in \eqref{eq:generating}. Minimum energy per bit and wideband slope are derived in the following result.
\begin{theo} \label{theo:fluid}
Assume that the source arrival is modeled by a two-state (ON-OFF) continuous-time Markov chain. Then, the minimum energy per bit and wideband slope expressions as a function of the QoS exponent $\theta$ are given, respectively, by
\begin{gather}
{\frac{E_b}{N_0}}_{\text{min}} = \frac{\log_e2}{\E\left\{ z\right\}} \label{eq:ebnomin-fluidtwostate-theo}
\intertext{and}
\hspace{-.4cm}\mathcal{S}_0=\frac{2(\E\left\{ z \right\})^2} {  \zeta \frac{\theta m}{\log_e2}(\E\left\{ z \right\})^2 +\frac{\theta}{m\log_e2}\sum_{i,j=1}^m \cov\left\{ z_i, z_j\right\} + \E\left\{ z^2 \right\} } \label{eq:widebandslope-fluidtwostate-theo}
\end{gather}
where $\zeta = \frac{2\beta}{\alpha (\alpha + \beta)}$ as defined in \eqref{eq:zeta}.
\end{theo}

\emph{Proof:} See Appendix \ref{subsec:fluidarrival}.

\begin{figure}
\begin{center}
\includegraphics[width=0.45\textwidth]{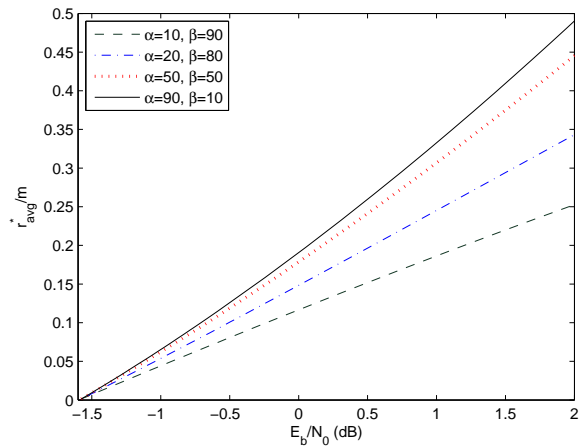}
\vspace{0.1cm}
\caption{Maximum average arrival rate $\frac{1}{m} r_{\avg}^*$ vs. energy per bit $\frac{E_b}{N_0}$ when $\rho=0.75$, $\theta = 1$, and $m=10$.}\label{fig:fluid_2}
\end{center}
\end{figure}

Similarly as before, QoS constraints and source randomness do not affect the minimum energy per bit. On the other hand, it is seen in  \eqref{eq:widebandslope-fluidtwostate-theo} that the impact of source arrival characteristics on the wideband slope is via the state transition rates $\alpha$ and $\beta$. For instance, larger the $\alpha$ value, the higher the wideband slope is.
This is due to the fact that  as $\alpha$, which is the transition rate from OFF state to ON state, increases, the system is more likely to be in the ON state. Contrarily, wideband slope diminishes with increasing $\beta$. This is expected as well since larger $\beta$ leads to higher OFF-state probabilities. The effect of $\alpha$ and $\beta$ is illustrated in Fig. \ref{fig:fluid_2}, where maximum average arrival rate vs. energy per bit is plotted. In this figure, we set $\theta = 1$ and $\rho = 0.75$. As predicted, the same minimum energy per bit is achieved for different values of $\alpha$ and $\beta$, while wideband slope increases with increasing $\alpha$ or decreasing $\beta$.

\subsubsection{n-State Markov Fluid Birth-Death Process}
In this subsection, we consider a birth-death process for the Markov fluid source. We assume that there are $n$ states and the arrival rate in the $i^{th}$ state is $(i-1)\lambda$ for $i = 1, \ldots, n$. The generating matrix for the birth-death process is in the form of
\begin{align}
\hspace{-0.4cm}\G=
\begin{bmatrix}
 -\alpha & \alpha & 0&\cdots&\cdots&0\\
 \beta&-(\alpha+\beta)&\alpha &0 & \cdots & 0 \\
 0 && \ddots &&& \vdots \\
 \vdots &&  &\ddots&& 0 \\
 0 & \cdots & 0 &\beta&-(\alpha+\beta)&\alpha \\
 0&\cdots&\cdots&0&-\beta&\beta \label{eq:birthdeathG}
\end{bmatrix}.
\end{align}
Hence, the transition rate from state $i$ to state $i+1$ is $\alpha$ whereas the transition rate from state $i$ to state $i-1$ is $\beta$. The effective bandwidth of this source, which does not have a simple closed-form expression, can be found from \eqref{eq:fluidebw}. In order to conduct an energy efficiency analysis, average arrival rate needs to be identified as well. Using \eqref{eq:stateprob} and \eqref{eq:birthdeathG}, we can easily determine that the stationary distribution as
\begin{align}
\pi_i=\frac{\left({\frac{\alpha}{\beta}}\right)^{i-1}-\left({\frac{\alpha}{\beta}}\right)^{i}}{1-\left({\frac{\alpha}{\beta}}\right)^n} \text{ for } i=1, 2, \ldots, n
\end{align}
when $\frac{\alpha}{\beta}\neq1$, $\alpha\neq0$ and $\beta\neq0$. If any of these inequalities is not satisfied, state probabilities can be obtained by limiting functions.

Now, under the assumptions that $\alpha\neq\beta$ and the arrival rate in state $i$ is $\lambda_i=(i-1)\lambda(\tsnr)$, the average arrival rate is given by
\begin{gather}
r_{\avg}=\frac{\xi(1-n\xi^{n-1}+(n-1)\xi^n)}{(1-\xi)(1-\xi^n)}\lambda(\tsnr) \label{eq:avgarrivalnstate}
\end{gather}
where $\xi=\tfrac{\alpha}{\beta}$.

\begin{rem}
Note that \eqref{eq:avgarrivalnstate} specializes to \eqref{eq:avgarrivaltwostate} if $n=2$. When $\xi\rightarrow\infty$, the probability of the $n^{th}$ state approaches $1$, and the arrival rate will be $(n-1)\lambda(\tsnr)$ at steady state. On the other hand, for $\xi=0$, the state of the source is stuck at the first state in which the arrival rate is zero.
\end{rem}

\begin{figure}
\begin{center}
\includegraphics[width=0.45\textwidth]{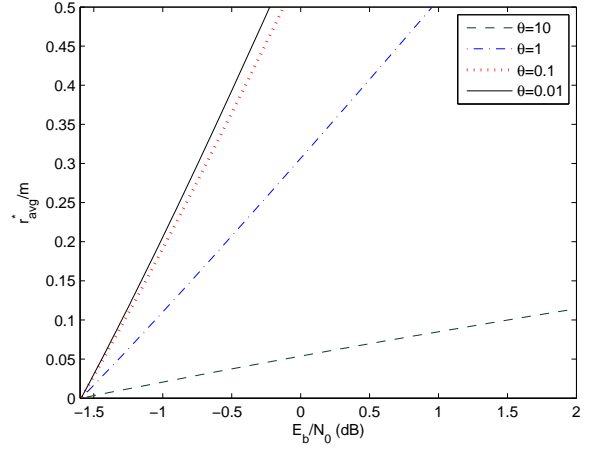}
\vspace{0.1cm}
\caption{Maximum average arrival rate $r_{\avg}^*$ vs. energy per bit $\frac{E_b}{N_0}$ when channel blocks are uncorrelated and $\alpha=\beta=50$. Number of states of the arrival model is $10$.}\label{fig:fluid_theta}
\end{center}
\end{figure}

Numerically, we can obtain the effective bandwidth of the $n$-state birth-death Markov fluid process using \eqref{eq:fluidebw}. Subsequently, solving \eqref{eq:equalityforQoS} and incorporating \eqref{eq:avgarrivalnstate}, we can determine the maximum average arrival rate $r_{\avg}^*(\tsnr)$, which we further employ for characterizing the energy efficiency. The results of this numerical analysis are displayed in the following figures. In Fig. \ref{fig:fluid_alpha},  we demonstrate the effect of $\alpha$ on the energy efficiency. In particular, when $\beta$ is kept fixed, increasing $\alpha$ improves the energy efficiency as in the two-state case.
\begin{figure}
\begin{center}
\includegraphics[width=0.45\textwidth]{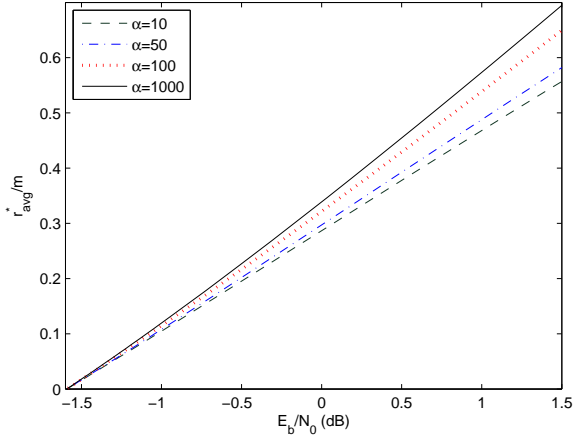}
\vspace{0.1cm}
\caption{Maximum average arrival rate $\frac{1}{m}r_{\avg}^*$ vs. energy per bit $\frac{E_b}{N_0}$ when channel blocks are uncorrelated. $\theta=1$, $\beta = 100$ and the number of states of the arrival process is $n = 10$.}\label{fig:fluid_alpha}
\end{center}
\end{figure}
We illustrate the effect of QoS constraints in Fig. \ref{fig:fluid_theta}. Similar conclusions as before readily apply. QoS exponent $\theta$ does not alter the minimum bit energy, which is $-1.59$ dB again, but the wideband slope is reduced with increasing $\theta$.

\subsection{Energy Efficiency with Markov-Modulated Poisson Process} \label{subsec:MMPP}

\subsubsection{ON-OFF Markov-Modulated Poisson Process}
Again, we initially address the two-state model in which there are no arrivals in the OFF state and the intensity of the Poisson arrival process is $\lambda$ in the ON state. The generating matrix $\G$ is the same as in \eqref{eq:generating}.
\begin{theo} \label{theo:MMPP}
Assume that the source arrival is modeled by a two-state (ON/OFF) Markov modulated Poisson process. Then, the minimum energy per bit and wideband slope expressions as a function of the QoS exponent $\theta$ are given, respectively, by
\begin{gather}
{\frac{E_b}{N_0}}_{\text{min}} = \frac{e^\theta-1}{\theta}\frac{\log_e2}{\E\left\{ z\right\}} \label{eq:ebnomin-poistwostate-theo}
\intertext{and}
\hspace{-.4cm}\mathcal{S}_0=\frac{\frac{2\theta}{e^\theta-1}(\E\left\{ z \right\})^2} {  \zeta \frac{\theta m}{\log_e2}(\E\left\{ z \right\})^2 +\frac{\theta}{m\log_e2}\sum_{i,j=1}^m \cov\left\{ z_i, z_j\right\} + \E\left\{ z^2 \right\} } \label{eq:widebandslope-poistwostate-theo}
\end{gather}
where $\zeta = \frac{2\beta}{\alpha (\alpha + \beta)}$ as defined in \eqref{eq:zeta}.
\end{theo}
\emph{Proof:} See Appendix \ref{subsec:MMPParrival}.
\begin{rem}
It is interesting to observe that, unlike the previous arrival models, minimum energy per bit in the case of MMPP source depends on the QoS exponent $\theta$. More specifically, minimum energy per bit increases with $(e^{\theta}-1)/\theta$ which is an increasing monotonic function of $\theta$ and always greater than one for $\theta>0$. On the other hand, as $\theta\to0$, $(e^{\theta}-1)/\theta\to1$.
Therefore, ${\frac{E_b}{N_0}}_{\text{min}} \ge \frac{\log_e2}{\E\left\{ z\right\}}$ with equality only if no QoS constraints are imposed (i.e., when $\theta = 0$).
Furthermore, in addition to its significant impact on the minimum energy per bit, increasing $\theta$ leads to much quicker reduction in the wideband slope due to the presence of the term $\frac{\theta}{e^{\theta}-1}$ in \eqref{eq:widebandslope-poistwostate-theo}. Hence, overall, energy costs grow very fast as $\theta$ increases. This is again because of the additional randomness arising from Poisson arrivals in the ON state.
\end{rem}
\begin{rem}
From \eqref{eq:widebandslope-poistwostate-theo}, we note that the effect of the state transition rates $\alpha$ and $\beta$ on the energy efficiency is the same as in the Markov fluid source model. Increasing $\alpha$ or decreasing $\beta$ improves the energy efficiency of the system because the burstiness of the data arrivals is reduced and the buffer overflows can be avoided at lower energy costs.
\end{rem}

\begin{figure}
\begin{center}
\includegraphics[width=0.45\textwidth]{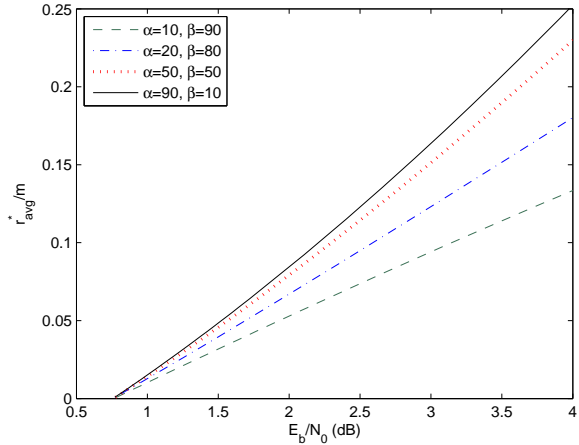}
\vspace{0.1cm}
\caption{Maximum average arrival rate $\frac{1}{m} r_{\avg}^*$ vs. energy per bit $\frac{E_b}{N_0}$ when $\rho=0.75$, $\theta = 1$, and $m=10$.}\label{fig:MMPP_2}
\end{center}
\end{figure}

We plot the maximum average arrival rate vs. energy per bit in Fig. \ref{fig:MMPP_2}. We set $\E\{z\}$ = 1 and $\theta=1$ for which the minimum energy per bit is $0.76$ dB. The increase in bit energy  with respect to $-1.59$ dB is due to $10\log_{10}((e^\theta-1)/\theta) = 10\log_{10}(e-1)$ for $\theta =1$. From the figure, we can again infer that adjusting $\alpha$ or $\beta$ to increase the ON state probability makes the system more energy efficient due to the increase in the wideband slope.

\subsubsection{n-State Markov-Modulated Poisson Process}
Finally, we consider an $n$-state MMPP process  and assume that the intensity of the Poisson arrivals in the $i^{th}$ state is $(i-1)\lambda$. For the Markov transitions between states, we consider the birth-death process and adopt the transition rate matrix $\G$ from \eqref{eq:birthdeathG}. We solve for the maximum intensity $\lambda^*(\tsnr, \theta)$ by incorporating \eqref{eq:poisEBW} into \eqref{eq:equalityforQoS}. Then, using the expression in \eqref{eq:avgarrivalnstate}, we obtain $r_\avg^*(\tsnr, \theta)$.

\begin{figure}
\begin{center}
\includegraphics[width=0.45\textwidth]{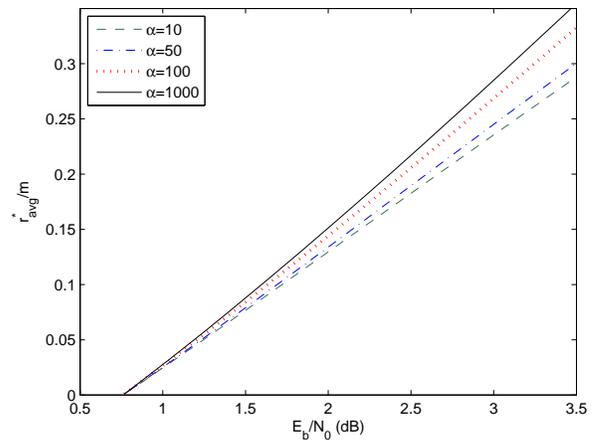}
\vspace{0.1cm}
\caption{Maximum average arrival rate $\frac{1}{m}r_{\avg}^*$ vs. energy per bit $\frac{E_b}{N_0}$ when channel blocks are uncorrelated. $\theta=1$, $\beta = 100$ and the number of Markov states for the arrival process is $n=10$.}\label{fig:MMPP_alpha}
\end{center}
\end{figure}

\begin{figure}
\begin{center}
\includegraphics[width=0.45\textwidth]{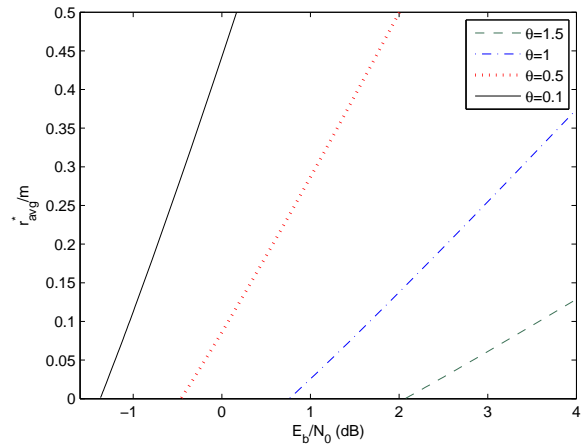}
\vspace{0.1cm}
\caption{Maximum average arrival rate $\frac{1}{m}r_{\avg}^*$ vs. energy per bit $\frac{E_b}{N_0}$ when channel blocks are uncorrelated and $\alpha=\beta=50$. Number of Markov states for the arrival process is $n = 10$.}\label{fig:MMPP_theta}
\end{center}
\end{figure}

In Figs. \ref{fig:MMPP_alpha} and \ref{fig:MMPP_theta}, we depict the maximum average arrival rate as a function of the energy per bit with uncorrelated channel coefficients being assumed in each block. In Fig. \ref{fig:MMPP_alpha},  we set $n=10$, $\theta=1$ and $\beta=100$, and demonstrate how $\alpha$ influences the energy efficiency of system. The observation has similarities with other Markovian sources regarding the source burstiness. Interestingly, the minimum energy per bit is again $0.76$ dB as in the two-state case, leading to the conclusion that the number of states does not alter $\frac{E_b}{N_0}_{\min}$ in this case. The degradation in energy efficiency due to increased $\theta$ is shown in Fig. \ref{fig:MMPP_theta}. As described in the two-state case, higher values of $\theta$ (i.e., stricter QoS constraints) result in higher $\frac{E_b}{N_0}_{\min}$ and smaller wideband slope. Therefore, even for relatively small increases in $\theta$, we can have large gaps between curves, indicating significantly high energy costs.

\section{Conclusion} \label{sec:conclusion}

In this paper, we have studied the throughput and energy efficiency in wireless fading channels when data arrivals to the buffer are random and constraints on buffer overflow probabilities are imposed. We have considered discrete-time Markov, Markov fluid, and Markov-modulated Poisson (MMPP) sources, and formulated the maximum average arrival rates of these sources in closed-form in the special case in which the source has only ON and OFF states. We have shown that as the QoS exponent $\theta$ vanishes (i.e., no QoS constraints are imposed), throughput converges to the ergodic capacity and the performance becomes independent of the source characteristics and channel fading correlations. On the other hand, we have demonstrated for $\theta > 0$ that the statistics of the source arrivals and channel conditions have significant impact on the throughput. More specifically, it is seen that while fast channel variations (e.g., having fading less correlated) can be beneficial, source randomness and burstiness degrade the throughput. In particular, performance loss can be severe with MMPP sources especially if $\theta$ is large. Similar conclusions are drawn in the energy efficiency analysis as well. We have shown that the minimum energy per bit is only a function of the marginal distribution of the fading for constant, discrete-time Markov and Markov fluid arrivals. However, for MMPP arrivals, this asymptotic measure of energy efficiency increases as the QoS exponent $\theta$ increases and hence as stricter queueing constraints are imposed. For practical scenarios, wideband slope is a better indicator of the performance and is always seen to depend on the source statistics, channel correlations, and QoS exponent $\theta$. In particular, we have noted that wideband slope diminishes with increasing source burstiness and stricter QoS constraints, indicating increased energy costs in such cases. While this paper primarily concentrates on two-state Markovian sources, numerical results are provided in the energy efficiency analysis for several simple $n$-state models with which we have similar observations as above.

\appendix

\subsection{Proof of Theorem \ref{theo:ravgdisctheo}:}\label{subsec:ravgdisc}

Using the effective bandwidth formulation in (\ref{eq:2discreteEBW}), we can express  \eqref{eq:equalityforQoS} in the following equivalent form:
\begin{align}
\frac{1}{\theta} \log_e\!\!\left(\!\!\tfrac{p_{11}+p_{22} e^{\lambda\theta}+\sqrt{ (p_{11}+p_{22}e^{\lambda\theta})^2 - 4(p_{11}+p_{22}-1)e^{\lambda\theta} }  }{2}\right)\! = C_E.
\end{align}
Then, we rewrite the above equality as
\begin{gather}
p_{11}\!+\!p_{22} e^{\lambda\theta}\! +\! \sqrt{ \!(p_{11}\!+p_{22}e^{\lambda\theta})^2\! - 4(p_{11}\!+p_{22}-\!1)e^{\lambda\theta} }\! = 2e^{\theta C_E},
\end{gather}
from which, after moving the first two terms on the left-hand side to the right-hand side and taking the square of both sides, we obtain
\begin{gather}
(p_{11}\!+p_{22}e^{\lambda\theta})^2\! - 4(p_{11}\!+p_{22}\!-\!1)e^{\lambda\theta} = \left(\!2e^{\theta C_E} - p_{11}\!-\!p_{22} e^{\lambda\theta}\!\right)^2 \!.
\end{gather}
Now, by simply exchanging the second term on the left-hand side with the term on the right-hand side, we have
\begin{align}
(p_{11}\!+p_{22}e^{\lambda\theta})^2\! - \left(\!2e^{\theta C_E} - p_{11}\!-\!p_{22} e^{\lambda\theta}\!\right)^2 \! &= 4(p_{11}\!+p_{22}\!-\!1)e^{\lambda\theta},
\\
\left(2p_{11}\!+2p_{22} e^{\lambda\theta}\!-2e^{\theta C_E}\right) 2e^{\theta C_E} &= 4(p_{11}\!+p_{22}-1)e^{\lambda\theta}.
\end{align}
After further rearrangements, we have
\begin{gather}
(p_{11}+p_{22}-1-p_{22}e^{\theta C_E})e^{\lambda\theta} = p_{11}e^{\theta C_E}-e^{2\theta C_E} .
\end{gather}
Solving the equation for $\lambda$, we get
\begin{align}
\lambda^*(\tsnr, \theta)=\frac{1}{\theta}\log_e\left(\frac{e^{2\theta C_E(\ssnr, \theta)} - p_{11}e^{\theta C_E(\ssnr, \theta)}} {1-p_{11}-p_{22}+p_{22}e^{\theta C_E(\ssnr, \theta)}}\right) \label{eq:2discreter}
\end{align}
which provides the maximum arrival rate in the ON state.
We can now express the maximum arrival rate as $r^*(\tsnr, \theta) = P_{\on} \lambda^*(\tsnr, \theta)$ and obtain the expression in \eqref{eq:2discreteravg}. \hfill $\Box$

\subsection{Proof of Theorem \ref{cor:discrete-Markov-theta-0}:}\label{subsec:discrete-Markov-theta-0}

Let us define
\begin{align}
\psi(\theta) = e^{-\theta C_E(\tsnr, \theta)} =\E\left\{ e^{-\tfrac{\theta}{\log_e2} \sum_{i=1}^{m} \log_e (1 + \ssnr z_i)}\right\}. \label{eq:psidef}
\end{align}
The following properties of $\psi$ can be verified easily:
\begin{align}
\psi(0) =& 1,
\\
\dot{\psi}(\theta) =&\E\left\{-\sum_{i = 1}^m \log_2(1 + \ssnr z_i) e^{-\theta \sum_{i = 1}^m \log_2(1 + \ssnr z_i)}\right\},
\\
\ddot{\psi}(\theta) =&\E\left\{\left(\sum_{i = 1}^m \log_2(1 + \ssnr z_i)\right)^2 e^{-\theta \sum_{i = 1}^m \log_2(1 + \ssnr z_i)}\right\},
\intertext{and}
\dot{\psi}(0) =& -\E\left\{\sum_{i = 1}^m \log_2(1 + \ssnr z_i)\right\} \label{eq:dot(psi)(0)},
\\
\ddot{\psi}(0) =& \E\left\{\left(\sum_{i = 1}^m \log_2(1 + \ssnr z_i)\right)^2\right\} \label{eq:ddot(psi)(0)}
\end{align}
where $\dot{\psi}$ and  $\ddot{\psi}$ denote the first and second derivatives of $\psi$ with respect to $\theta$, respectively. Additionally, we define $r_\avg^*(\tsnr, \theta)$ as
\begin{equation}
r_\avg^*(\tsnr, \theta)=\frac{f_1(\theta)}{\theta}.
\end{equation}
Therefore, by applying L'Hopital's rule and letting $\theta \to 0$, maximum average arrival rate and its slope can be easily found as
\begin{align}
\lim_{\theta \to 0}\! r_\avg^*(\tsnr, \theta)\! &= \dot {f_1}(0), \label{eq:fravgtheta0disc}
\\
\left.\frac{\partial r_\avg^*(\tsnr, \theta)}{\partial\theta}\right|_{\theta=0}&= \frac{\ddot {f_1}(0)}{2}. \label{eq:fdotravgtheta0disc}
\end{align}
Now, replacing $e^{-\theta C_E(\tsnr, \theta)}$ with $\psi(\theta)$ in the expression of $r_\avg^*(\tsnr, \theta)$ in \eqref{eq:2discreteravg}, we can express $f_1(\theta)$ as
\small{
\begin{equation}
\hspace{-0.1cm}f_1(\!\theta)\!=\!P_{\on}\!\left[\log_e\!\left(1\!-\!p_{11}\psi(\theta)\right)\!-\!\log_e\!\left(\left(1\!-\!p_{11}\!-\!p_{22}\right)\!\psi^2(\theta)  \!+\!p_{22}\psi(\theta)\right)\!\right]. \label{eq:fthetadisc}
\end{equation}}
\normalsize
Therefore, we derive $\dot {f_1}(\theta) $ when $\theta\to0$ as
\begin{align}
\dot {f_1}(0) &= P_{\on}\! \left[\frac{-p_{11} \dot{\psi}(0)}{1\!-\!p_{11}\psi(0)}-\frac{(1\!-\!p_{11}\!-\!p_{22})\dot{\psi}(0)}{(1\!-\!p_{11}\!-\!p_{22})\psi(0)\!+\!p_{22}} \!-\!\frac{\dot{\psi}(0)}{\psi(0)} \right] \label{eq:limr_avg_theta_0_1}
\\
&= P_{\on}\! \left[\frac{-p_{11}}{1\!-\!p_{11}}-\frac{1\!-\!p_{11}\!-\!p_{22}}{1\!-\!p_{11}}-1 \right] \dot{\psi}(0) \label{eq:limr_avg_theta_0_3}
%
\\
&=\E\left\{\sum_{i = 1}^m \log_2(1 + \ssnr z_i)\right\}. \label{eq:limr_avg_theta_0_4}
\end{align}
Note that \eqref{eq:limr_avg_theta_0_1} follows by taking the first derivative of the expression in \eqref{eq:fthetadisc} with respect to $\theta$, and (\ref{eq:limr_avg_theta_0_3}) is obtained using the property that $\psi(0) =1$. Finally, (\ref{eq:limr_avg_theta_0_4}) and hence the result in \eqref{eq:2discreteravg-theta-0} immediately follow from (\ref{eq:P_ON}), \eqref{eq:dot(psi)(0)} and \eqref{eq:fravgtheta0disc}. 
\begin{figure*}
\begin{align}
\ddot {f_1}(0) &= P_{\on}\frac{d}{d\theta}\! \left.\left[\frac{-p_{11} \dot{\psi}(0)}{1\!-\!p_{11}\psi(0)}-\frac{(1\!-\!p_{11}\!-\!p_{22})\dot{\psi}(0)}{(1\!-\!p_{11}\!-\!p_{22})\psi(0)\!+\!p_{22}} \!-\!\frac{\dot{\psi}(0)}{\psi(0)} \right]\right|_{\theta=0} \label{eq:discdotravgtheta0_1}
\\
&= P_{\on}\! \left\{\left[\frac{-p_{11}}{1\!-\!p_{11}}-\frac{1\!-\!p_{11}\!-\!p_{22}}{1\!-\!p_{11}}-\!1 \right] \ddot{\psi}(0)+ \!\left[-\frac{p_{11}^2}{(1\!-\!p_{11})^2} +\frac{(1\!-\!p_{11}\!-\!p_{22})^2}{(1\!-\!p_{11})^2}\!+\!1  \right]\left(\dot{\psi}(0)\right)^2\right\} \label{eq:discdotravgtheta0_2}
\\
&=-\ddot{\psi}(0)+(1-\eta)\left(\dot{\psi}(0)\right)^2 \label{eq:discdotravgtheta0_3}
\\
&=-\E\left\{\left(\sum_{i = 1}^m \log_2(1 + \ssnr z_i)\right)^2\right\}+(1-\eta)\left[\E\left\{\sum_{i = 1}^m \log_2(1 + \ssnr z_i)\right\}\right]^2 \label{eq:discdotravgtheta0_4}
\end{align}
\end{figure*}

Next, we determine the slope of the throughput in \eqref{eq:dotravgdisctheta0} as the QoS exponent $\theta$ approaches zero. For this, we only need to derive the second derivative expression $\ddot f_1(0)$, which is done at the top of the next page. \eqref{eq:discdotravgtheta0_1}, \eqref{eq:discdotravgtheta0_2} and \eqref{eq:discdotravgtheta0_3} follow from straightforward algebraic steps. Inserting \eqref{eq:dot(psi)(0)} and \eqref{eq:ddot(psi)(0)} into \eqref{eq:discdotravgtheta0_3}, we obtain \eqref{eq:discdotravgtheta0_4}. Finally, the result in \eqref{eq:dotravgdisctheta0} follows by combining \eqref{eq:discdotravgtheta0_4} and \eqref{eq:fdotravgtheta0disc}. \hfill $\square$

\subsection{Proof of Theorem \ref{cor:disc-high-SNR}:}\label{subsec:disc-high-SNR}

In the analysis of the high-SNR slope of the effective capacity, it has been shown in \cite{gursoy-mimo} that
\begin{align}
-\frac{1}{\theta} &\log_e \E\left\{e^{-\theta \log_2 (1 + \tsnr z)}\right\} \nonumber
\\
&=
\left\{
\begin{array}{ll}
\frac{1}{\theta \log_2e} \log_2 \tsnr + \mathcal{O}(1) & \text{if } \theta > \frac{1}{\log_2e}
\\
\log_2 \tsnr + \mathcal{O}(1) &\text{if } 0 < \theta < \frac{1}{\log_2e}
\end{array}\right. \label{eq:asymptotic-mimo-paper}
\end{align}
where $z$ is exponentially distributed with unit mean. If we assume that fading in each block is i.i.d., then the effective capacity expression in (\ref{eq:effcap}) becomes
\begin{align}
C_E(\tsnr,\theta) &= -\frac{1}{\theta}\log_e\E \left\{e^{-\theta
\sum_{i = 1}^{m} \log(1 + \ssnr z_i)}\right\}
\\
&= -\frac{1}{\theta}\log_e \left(\prod_{i = 1}^m \E \left\{e^{-\theta
\log(1 + \ssnr z_i)}\right\}\right)
\\
&= -\frac{m}{\theta}\log_e \E \left\{e^{-\theta
\log(1 + \ssnr z)}\right\}. \label{eq:effcap-iid-case}
\end{align}
Furthermore, the maximum average arrival rate in (\ref{eq:2discreteravg}) can be expressed as
\begin{align}
&r_\avg^*(\tsnr, \theta) \nonumber
\\
&=\frac{P_{\on}}{\theta}\log_e \!\!\left(\! \frac{e^{2\theta C_E(\ssnr, \theta)}\left(1 - p_{11}e^{-\theta C_E(\ssnr, \theta)}\right)} {e^{\theta C_E(\ssnr, \theta)} \left(\left(1-p_{11}-p_{22}\right) e^{-\theta C_E(\ssnr, \theta)} +p_{22}\right)}\!\right)
\\
&=\frac{P_{\on}}{\theta}\log_e \left( \frac{e^{\theta C_E(\ssnr, \theta)}\left(1 - p_{11}e^{-\theta C_E(\ssnr, \theta)}\right)} {\left(1-p_{11}-p_{22}\right) e^{-\theta C_E(\ssnr, \theta)} +p_{22}}\right) \label{eq:ravg-high-SNR-analysis1}
\\
&=\frac{P_{\on}}{\theta} \bigg(\log_e e^{\theta C_E(\ssnr, \theta)}
+\log_e \left(1 - p_{11}e^{-\theta C_E(\ssnr, \theta)}\right) \nonumber
\\
&\hspace{1.3cm}-\log_e \left(\left(1-p_{11}-p_{22}\right) e^{-\theta C_E(\ssnr, \theta)} +p_{22}\right)\bigg)\label{eq:ravg-high-SNR-analysis2}
\\
&= \frac{P_{\on}}{\theta} \log_e e^{\theta C_E(\ssnr, \theta)} + \mathcal{O}(1) \label{eq:ravg-high-SNR-analysis3}
\\
&= P_{\on} \, C_E(\ssnr, \theta) + \mathcal{O}(1) \label{eq:ravg-high-SNR-analysis4}
\end{align}
where (\ref{eq:ravg-high-SNR-analysis1}) and (\ref{eq:ravg-high-SNR-analysis2}) follow from straightforward algebraic operations and \eqref{eq:ravg-high-SNR-analysis3} is due to the fact that $C_E(\ssnr, \theta)$ increases without bound as $\tsnr$ increases and hence the term $e^{-\theta C_E(\ssnr, \theta)}$ vanishes asymptotically in the formulations.

Finally, combining \eqref{eq:asymptotic-mimo-paper}, \eqref{eq:effcap-iid-case}, and \eqref{eq:ravg-high-SNR-analysis4}, we immediately obtain the desired result in (\ref{eq:high-SNR-expansion-discretemarkov}) for the cases in which $\theta > 0$. When $\theta = 0$, the result follows from \eqref{eq:2discreteravg-theta-0} in Theorem \ref{cor:discrete-Markov-theta-0}. \hfill $\Box$

\subsection{Proof of Theorem \ref{theo:ravgfluidtheo}:}\label{subsec:ravgfluid}

Using (\ref{eq:2fluidEBW}), we can rewrite \eqref{eq:equalityforQoS} as
\begin{gather}
\left(\theta \lambda -(\alpha+\beta)\!-2\theta C_E\right)^2 \!=(\theta  \lambda -(\alpha+\beta))^2+4\alpha\theta \lambda \label{eq:equalityfluid}
\end{gather}
which can further be simplified to
\begin{align}
-2\theta C_E\!\left(2\theta \lambda -2(\alpha+\beta)\!-2\theta C_E\right)  &= 4\alpha\theta \lambda
.\label{eq:equalityfluid2}
\end{align}
Next, solving for $\lambda$, we obtain
\begin{align}
 \lambda^*(\tsnr, \theta)=\frac{\theta C_E(\ssnr, \theta) +\alpha+\beta} {\theta C_E(\ssnr, \theta) +\alpha} \, C_E(\tsnr, \theta). \label{eq:2fluidr}
\end{align}
Finally, using the expression in \eqref{eq:avgarrivaltwostate}, we derive the maximum average arrival rate given in \eqref{eq:2fluidravg}. \hfill $\Box$

\subsection{Proof of Theorem \ref{cor:fluid-Markov-theta-0}:}\label{subsec:fluid-Markov-theta-0}

Similar as in the Proof of Theorem \ref{cor:discrete-Markov-theta-0} in Appendix \ref{subsec:discrete-Markov-theta-0}, we define $r_\avg^*(\tsnr, \theta)=\frac{f_2(\theta)}{\theta}$ with
\begin{equation}
f_2(\theta)=-P_{\on} \frac{\alpha+\beta-\log_e\psi(\theta)}{\alpha-\log_e\psi(\theta)} \log_e\psi(\theta). \label{eq:fthetafluid}
\end{equation}
Now, we have \eqref{eq:fravgtheta0disc} and \eqref{eq:fdotravgtheta0disc} hold with $f_1$ replaced with $f_2$.
The remainder of the proof requires only the determination of the first and second derivatives of $f_2(\theta)$ at $\theta = 0$. The first derivative $\dot {f_2}(0)$ is given at the top of the next page in \eqref{eq:limr_avg_theta_0_1f}-\eqref{eq:limr_avg_theta_0_4f}. Note that \eqref{eq:limr_avg_theta_0_1f} and \eqref{eq:limr_avg_theta_0_2f} follow from straightforward algebraic steps, and \eqref{eq:limr_avg_theta_0_3f} is obtained by noting the property that $\psi(0) =1$. Finally, (\ref{eq:limr_avg_theta_0_4f}) and hence the result in \eqref{eq:2fluidravg-theta-0} immediately follow from (\ref{eq:Pontwostate}), \eqref{eq:dot(psi)(0)} and \eqref{eq:fravgtheta0disc}.
\begin{figure*}
\begin{align}
\dot {f_2}(0) &= \lim_{\theta \to 0} -P_{\on}\!  \left\{ \left[\frac{-\frac{\dot\psi(\theta)}{\psi(\theta)}}{\alpha-\log_e\psi(\theta)}- \frac{(\alpha+\beta-\log_e\psi(\theta))\left(-\frac{\dot\psi(\theta)}{\psi(\theta)}\right)}{(\alpha-\log_e\psi(\theta))^2} \right]\log_e\psi(\theta) + \frac{\alpha+\beta-\log_e\psi(\theta)}{\alpha-\log_e\psi(\theta)} \frac{\dot\psi(\theta)}{\psi(\theta)}\right\}   \label{eq:limr_avg_theta_0_1f}
\\
&= \lim_{\theta \to 0} -P_{\on}\!  \left[ 1 + \frac{\alpha\beta}{(\alpha-\log_e\psi(\theta))^2} \right]\frac{\dot\psi(\theta)}{\psi(\theta)} \label{eq:limr_avg_theta_0_2f}
\\
&= -P_{\on}\! \frac{\alpha+\beta}{\alpha} \dot{\psi}(0) \label{eq:limr_avg_theta_0_3f}
%
\\
&=\E\left\{\sum_{i = 1}^m \log_2(1 + \ssnr z_i)\right\} \label{eq:limr_avg_theta_0_4f}
\end{align}
\end{figure*}

Next, we obtain the slope expression in \eqref{eq:dotravgfluidtheta0} in the limit as the QoS exponent $\theta$ approaches zero. For this, we characterize the second derivative expression $\ddot f_2(0)$ on the next page in \eqref{eq:fluiddotravgtheta0_1}--\eqref{eq:fluiddotravgtheta0_4} . \eqref{eq:fluiddotravgtheta0_1}, \eqref{eq:fluiddotravgtheta0_2} are readily obtained and \eqref{eq:fluiddotravgtheta0_3} is determined by noting that $\psi(0) =1$. We incorporate \eqref{eq:dot(psi)(0)} and \eqref{eq:ddot(psi)(0)} into \eqref{eq:fluiddotravgtheta0_3} to obtain \eqref{eq:fluiddotravgtheta0_4}. The result in \eqref{eq:dotravgfluidtheta0} follows by combining \eqref{eq:fluiddotravgtheta0_4} and \eqref{eq:fdotravgtheta0disc} (with $f_1$ replaced with $f_2$). \hfill $\square$
\begin{figure*}
\begin{align}
\ddot {f_2}(0) &= \lim_{\theta \to 0} -P_{\on}\!  \frac{d}{d\theta}\left\{\left[ 1 + \frac{\alpha\beta}{(\alpha-\log_e\psi(\theta))^2} \right]\frac{\dot\psi(\theta)}{\psi(\theta)} \right\} \label{eq:fluiddotravgtheta0_1}
\\
&= \lim_{\theta \to 0} -P_{\on}\!  \frac{d}{d\theta}\left\{ \frac{-2\alpha\beta\left(-\frac{\dot\psi(\theta)}{\psi(\theta)}\right)} {(\alpha-\log_e\psi(\theta))^3}\left(\frac{\dot\psi(\theta)}{\psi(\theta)}\right)+ \left[ 1 + \frac{\alpha\beta}{(\alpha-\log_e\psi(\theta))^2} \right]\left[\frac{\ddot\psi(\theta)}{\psi(\theta)}-\left(\frac{\dot\psi(\theta)}{\psi(\theta)}\right)^2\right]\right\} \label{eq:fluiddotravgtheta0_2}
\\
&=-\ddot{\psi}(0)+(1-\frac{2\beta}{\alpha(\alpha+\beta)})\left(\dot{\psi}(0)\right)^2 \label{eq:fluiddotravgtheta0_3}
\\
&=-\E\left\{\left(\sum_{i = 1}^m \log_2(1 + \ssnr z_i)\right)^2\right\}+ (1-\frac{2\beta}{\alpha(\alpha+\beta)})\left[\E\left\{\sum_{i = 1}^m \log_2(1 + \ssnr z_i)\right\}\right]^2 \label{eq:fluiddotravgtheta0_4}
\end{align}
\end{figure*}

\subsection{Proof of Theorem \ref{theo:ravgMMPPtheo}:}\label{subsec:ravgMMPP}
We find the maximum average arrival rate $r_\avg^*(\tsnr, \theta)$ by incorporating \eqref{eq:2MMPPEBW} into (\ref{eq:equalityforQoS}) and expressing \eqref{eq:equalityforQoS} as
\begin{align}
&\left((e^\theta-1)  \lambda -(\alpha+\beta)-2\theta C_E\right)^2 \nonumber
\\
&\hspace{1cm}=\left((e^\theta-1)  \lambda -(\alpha+\beta)\right)^2+4\alpha(e^\theta-1)  \lambda.
\end{align}
Similarly as in the proof of Theorem \ref{theo:ravgfluidtheo}, we can simplify the above equality and solve for the maximum Poisson arrival intensity in the ON state to obtain
\begin{gather}
 \lambda^*(\tsnr, \theta)=\frac{\theta\left[\theta C_E(\ssnr, \theta) +\alpha+\beta\right]} {(e^\theta\!\!-\!1)\left[\theta C_E(\ssnr, \theta) +\alpha\right]} \, C_E(\tsnr, \theta). \label{eq:2MMPPr}
\end{gather}
With this characterization, the maximum average arrival rate is readily obtained from \eqref{eq:poisavgarrivaltwostate}. \hfill $\Box$

\subsection{Proof of Theorem \ref{cor:MMPP-Markov-theta-0}:}\label{subsec:MMPP-Markov-theta-0}

Employing $f_2(\theta)$ defined in \eqref{eq:fthetafluid}, we can express the maximum average arrival rate as
\begin{equation}
r_\avg^*(\tsnr, \theta)\!= \frac{f_2(\theta)}{e^\theta-1}. \label{eq:fthetaMMPP}
\end{equation}

Then, the throughput in the limit as $\theta$ approaches zero is given by
\begin{equation}
\lim_{\theta \to 0}\! r_\avg^*(\tsnr, \theta)\! = \lim_{\theta \to 0} \frac{\dot {f_2}(\theta)}{e^\theta} =\dot {f_2}(0). \label{eq:fravgtheta0MMPP}
\end{equation}
Inserting the result from \eqref{eq:limr_avg_theta_0_4f} into \eqref{eq:fravgtheta0MMPP}, we obtain \eqref{eq:2MMPPravg-theta-0}. 
Next, we determine the slope of the throughput when $\theta$ approaches zero:
\begin{align}
\left.\frac{\partial \!r_\avg^*(\tsnr, \theta)}{\partial\theta}\right|_{\theta=0}&= \lim_{\theta \to 0} \frac{\dot {f_2}(\theta)}{e^\theta-1} -\frac{e^\theta f_2(\theta)}{\left(e^\theta-1\right)^2} \label{eq:MMPPdotravgtheta0_1}
\\
&= \lim_{\theta \to 0}\!\frac{\left(e^\theta-1\right)\dot {f_2}(\theta)-e^\theta f_2(\theta)}{\left(e^\theta-1\right)^2} \label{eq:MMPPdotravgtheta0_2}
\\
&=\lim_{\theta \to 0}\!\frac{\left(e^\theta-1\right)\ddot {f_2}(\theta)-e^\theta f_2(\theta)}{2\left(e^\theta-1\right)e^\theta} \label{eq:MMPPdotravgtheta0_3}
\\
&=\frac{\ddot {f_2}(0)}{2}-\frac{1}{2}\lim_{\theta \to 0}\frac{f_2(\theta)}{e^\theta-1} \label{eq:MMPPdotravgtheta0_4}
\\
&=\frac{\ddot {f_2}(0)}{2}-\frac{\dot {f_2}(0)}{2}. \label{eq:MMPPdotravgtheta0_5}
\end{align}
\eqref{eq:MMPPdotravgtheta0_1} follows by taking the derivative of the expression in \eqref{eq:fthetaMMPP} with respect to $\theta$. \eqref{eq:MMPPdotravgtheta0_2} is obtained by simplifying \eqref{eq:MMPPdotravgtheta0_1}. We apply L'Hopital's rule on \eqref{eq:MMPPdotravgtheta0_2} to get \eqref{eq:MMPPdotravgtheta0_3} and further simplify it in \eqref{eq:MMPPdotravgtheta0_4}. Finally, we obtain \eqref{eq:MMPPdotravgtheta0_5}, which we used to derive \eqref{eq:dotMMPPfluidtheta0} by inserting \eqref{eq:limr_avg_theta_0_4f} and \eqref{eq:fluiddotravgtheta0_4} into \eqref{eq:MMPPdotravgtheta0_5}.

\subsection{Proof of Theorem \ref{theo:correlatedfading}:}\label{subsec:constantarrival}
When the arrival rate is fixed, the following equality holds:
\begin{gather}
r_\avg^*(\tsnr,\theta)=C_E(\tsnr, \theta).
\end{gather}
Therefore, in formulas \eqref{eq:ebnomin-ra}, \eqref{eq:widebandslope-ra}, we can use $\dot\C_E(0)$ and $\ddot\C_E(0)$  instead of $\dot{r}_\avg^*(0)/m$ and $\ddot{r}_\avg^*(0)/m$ respectively, where we have defined $\C_E(\tsnr, \theta) = C_E(\tsnr,\theta)/m$ as the normalized effective capacity. Minimum energy per bit and wideband slope becomes
\begin{align}\label{ebmin}
\frac{E_b}{N_0}_{\tmin}=\frac{1}{\dot{\C}_{E}(0)},
\end{align}
and
\begin{equation}\label{slope}
\mathcal{S}_0=-\frac{2(\dot{\C}_E(0))^2}{\ddot{\C}_E(0)}\log_e{2}.
\end{equation}
Thus, we only need to obtain the first and second derivatives of $\C_E(\tsnr,\theta)$ with respect to $\tsnr$ at $\tsnr = 0$ to determine the minimum energy per bit and wideband slope. We first express the effective capacity given in (\ref{eq:effcap}) as
\begin{align}
\C_E(\tsnr) &= -\frac{1}{\theta m} \log_e \g(\tsnr)
\end{align}
where we have defined
\begin{gather}
\g(\tsnr) = \E\left\{ e^{-\tfrac{\theta}{\log_e2} \sum_{i=1}^{m} \log_e (1 + \ssnr z_i)}\right\}. \label{eq:gdef}
\end{gather}
Now, the first and second derivatives of $\C_E(\tsnr)$ with respect to SNR are easily seen to be given by
\begin{gather}
\dot\C_E(\tsnr) = -\frac{1}{\theta m}\frac{\dot \g(\tsnr)}{\g(\tsnr)}, \text{ and}
\\
\ddot\C_E(\tsnr) = -\frac{1}{\theta m} \frac{\ddot \g(\tsnr) \g(\tsnr) - [\dot \g(\tsnr)]^2}{[\g(\tsnr)]^2},
\end{gather}
where $\dot{\g}$ and $\ddot{\g}$ denote the first and second derivatives of the function $\g$ with respect to $\tsnr$ and can be expressed as
\begin{gather}
\dot{\g}(\tsnr)\!= -\frac{\theta}{\log_e2}\E\left\{\sum_{i=1}^m \frac{z_i}{1+\ssnr z_i} \, e^{-\frac{\theta}{m\log_e2} \sum_{i=1}^{m} \log_e (1 + \ssnr z_i)}\right\}
\end{gather}
and
\begin{align}
&\ddot{\g}(\tsnr) \nonumber
\\
&= \frac{\theta}{\log_e\!2} \E\Bigg\{\!\!\!\Bigg(\sum_{i=1}^m\! \frac{z_i^2}{(1\!+\!\ssnr z_i)^2} \!+\! \frac{\theta}{\log_e\!2} \sum_{i,j=1}^m\! \frac{z_i z_j}{(1\!+\!\ssnr z_i) (1\!+\!\ssnr z_j)}\!\Bigg) \nonumber
\\
&\hspace{1.4cm}\times e^{-\tfrac{\theta}{\log_e\!2} \sum_{i=1}^{m} \log_e (1 + \ssnr z_i)} \Bigg\}.
\end{align}
Then, at $\tsnr = 0$, we have
\begin{gather}
\dot\C_E(0) = \frac{ \sum_{i=1}^{m}\E\left\{ z_i\right\} }{ m\log_e2} = \frac{\E\{z\}}{\log_e2} \label{eq:firstderiv-snr0}
\end{gather}
and
\begin{align}
\ddot\C_E(0) = -\frac{\theta \sum_{i=1}^{m}\sum_{j=1}^{m} \cov\left\{ z_i, z_j\right\} + \log_e\!2\sum_{i=1}^{m}\E\left\{ z_i^2\right\}} {m (\log_e2)^2} \nonumber
\\
= -\frac{\theta \sum_{i,j=1}^{m}\sum_{j=1}^{m} \cov\left\{ z_i, z_j\right\} + m\log_e\!2\E\left\{ z^2\right\}} {m (\log_e2)^2} \label{eq:secondderiv-snr0}
\end{align}
where we have used the facts that $\sum_{i=1}^{m}\E\left\{ z_i\right\} = m \E\{z\}$ and $\sum_{i=1}^{m}\E\left\{ z_i^2\right\} = m \E\{z^2\}$ due to our assumption that the fading coefficients and therefore $\{z_i\}$'s are identically distributed.

Plugging the expressions in (\ref{eq:firstderiv-snr0}) and (\ref{eq:secondderiv-snr0}) into those in (\ref{ebmin}) and (\ref{slope}), we readily obtain the minimum energy per bit and wideband slope expressions in (\ref{eq:ebno-min}) and (\ref{eq:widebandslope}). \hfill $\blacksquare$

\subsection{Proof of Theorem \ref{theo:discrete}:}\label{subsec:discretearrival}
To show the result, we need to obtain the first and second derivatives of $r_\avg^*(\tsnr)$. We first express the maximum average arrival rate in \eqref{eq:2discreteravg} as
\begin{align}
r_\avg^*(\tsnr,\theta)=\frac{P_{\on}}{\theta}&\Big[\log_e(1-p_{11}\g(\tsnr)) -\log_e(\g(\tsnr))\Big. \nonumber
\\
&\Big.-\log_e\big((1-p_{11}-p_{22})\g(\tsnr)+p_{22}\big)\Big]
\end{align}
where we have used the definition that $e^{\theta C_E(\ssnr, \theta) }=\tfrac{1}{\g(\tsnr)}$ with $\g(\tsnr)$ defined in \eqref{eq:gdef}. Taking the first derivative with respect to $\tsnr$, we obtain
\begin{align}
\dot{r}_\avg^*(\tsnr,\theta)=\frac{P_{\on}}{\theta}& \Bigg[\frac{-p_{11}\dot{\g}(\tsnr)}{1-p_{11}\g(\tsnr)} -\frac{\dot{\g}(\tsnr)}{\g(\tsnr)}\Big. \nonumber
\\
&\Big.-\frac{(1-p_{11}-p_{22})\dot{\g}(\tsnr)}{(1-p_{11}-p_{22})\g(\tsnr)+p_{22}}\Bigg]. \label{eq:dotravg_disc}
\end{align}
Next, we let $\tsnr \to 0$. Since the arrival rate $\lambda \rightarrow 0$ when $\tsnr\rightarrow0$, the equality in \eqref{eq:dotravg_disc} becomes
\begin{align}
\dot{r}_\avg^*(0,\theta)&=\frac{\dot{\g}(0)}{\theta} P_{\on}\Bigg[-\frac{p_{11}}{1-p_{11}} -1-\frac{1-p_{11}-p_{22}}{1-p_{11}} \Bigg]
\\
&=-\frac{\dot \g(0)}{\theta}=\frac{1}{\log_e2}\sum_{i=1}^m \E \left\{  z_i \right\} = \frac{m\E\{z\}}{\log_e2} \label{eq:dotravg0_disc}
\end{align}
where $P_{\on}=\tfrac{1-p_{11}}{2-p_{11}-p_{22}}$. Plugging the result in \eqref{eq:dotravg0_disc} into (\ref{eq:ebnomin-ra}), we immediately obtain (\ref{eq:ebnomin-ra-theo}).

In order to find the wideband slope, we first determine the second derivative of the maximum average arrival rate with respect to $\tsnr$ and then evaluate it at $\tsnr = 0$ as follows:
\begin{align}
\ddot{r}_\avg^*(0,\theta)=&\frac{\ddot{\g}(0)}{\theta} P_{\on}\Bigg[-\frac{p_{11}}{1-p_{11}} -1-\frac{(1-p_{11}-p_{22})}{1-p_{11}} \Bigg] \nonumber
\\
&+\frac{\left[\dot{\g}(0)\right]^2}{\theta}P_{\on}\Bigg[-\frac{p_{11}^2}{(1-p_{11})^2} +1\!+\frac{(1-p_{11}-p_{22})^2}{(1-p_{11})^2} \Bigg]\nonumber
\\
=&-\frac{\ddot{\g}(0)}{\theta}+(1-\eta)\frac{\left[\dot{\g}(0)\right]^2}{\theta} \label{eq:ddotravg0_disc}.
\end{align}
(\ref{eq:ddotravg0_disc}) follows from the fact that $g(0) = 1$, and $\eta$ is defined in (\ref{eq:eta}). Finally, inserting \eqref{eq:dotravg0_disc} and \eqref{eq:ddotravg0_disc} into \eqref{eq:widebandslope-ra}, the wideband slope expression in \eqref{eq:widebandslope-ra-theo} is readily obtained.
\hfill $\blacksquare$

\subsection{Proof of Theorem \ref{theo:fluid}:}\label{subsec:fluidarrival}
We differentiate the maximum average arrival rate expression in \eqref{eq:2fluidravg} with respect to SNR and obtain
\begin{align}
\dot r_\avg^*(\tsnr, \theta)=P_{\on} \left\{ \frac{2 \theta \dot C_E(\tsnr)C_E(\tsnr) +(\alpha+\beta)\dot C_E(\tsnr)} {\theta C_E(\tsnr) +\alpha} \right. \nonumber
\\
\left.- \frac{\left[\theta C_E^2(\tsnr) +(\alpha+\beta) C_E(\tsnr)\right]\theta\dot C_E(\tsnr)} {(\theta C_E(\tsnr) +\alpha)^2} \right\}. \label{eq:dotravg_fluid}
\end{align}
As $\tsnr\rightarrow0$, we can easily derive
\begin{align}
\dot r_\avg^*(0, \theta)=P_{\on} \frac{\alpha+\beta}{\alpha}  \dot C_E(0)=  \dot C_E(0)=\frac{m\E\{z\}}{\log_e2} \label{eq:dotravg0fluid}
\end{align}
where we use the facts that $C_E(0)=0$ and $P_{\on}=\frac{\alpha}{\alpha+\beta}$. Plugging the result in (\ref{eq:dotravg0fluid}) into (\ref{eq:ebnomin-ra}), we immediately obtain (\ref{eq:ebnomin-fluidtwostate-theo}).

In order to determine the wideband slope, we additionally take the second derivative of the maximum average arrival rate with respect to $\tsnr$ and evaluate it at $\tsnr = 0$ as
\begin{align}
\ddot r_\avg^*(0, \theta)=\ddot C_E(0)-\frac{2\theta\beta}{\alpha(\alpha+\beta)} \left(\dot C_E(0)\right)^2. \label{eq:ddotravg0_fluid}
\end{align}
Now, inserting the results in \eqref{eq:dotravg0fluid} and \eqref{eq:ddotravg0_fluid} into \eqref{eq:widebandslope-ra} and using the formulations in \eqref{eq:firstderiv-snr0} and \eqref{eq:secondderiv-snr0}, we obtain \eqref{eq:widebandslope-fluidtwostate-theo}. \hfill $\blacksquare$

\subsection{Proof of Theorem \ref{theo:MMPP}:} \label{subsec:MMPParrival}
The proof is rather straightforward after realizing that $r_\avg^*(\tsnr, \theta)$ of the MMPP source given in \eqref{eq:2MMPPravg} is equal to the maximum average arrival rate of the Markov fluid source in \eqref{eq:2fluidravg} scaled with $\frac{\theta}{e^\theta - 1}$. Therefore, making use of the results in \eqref{eq:dotravg0fluid} and \eqref{eq:ddotravg0_fluid}, we can immediately express the first and second derivatives of $r_\avg^*(\tsnr, \theta)$ at $\tsnr = 0$ as
\begin{align}
\dot r_\avg^*(0, \theta)=  \frac{\theta\dot C_E(0)}{\left(e^\theta\!-1\right)}=\frac{\theta m\E\{z\}}{\left(e^\theta\!-1\right)\log_e2} \, , \label{eq:dotRavg0MMPP}
\end{align}
\begin{align}
\ddot r_\avg^*(0, \theta)=\frac{\theta}{\left(e^\theta\!-1\right)}\left[\ddot C_E(0)-\frac{2\theta\beta}{\alpha(\alpha+\beta)} \left(\dot C_E(0)\right)^2\right]. \label{eq:ddotRavg0MMPP}
\end{align}
Then, the expressions in \eqref{eq:ebnomin-poistwostate-theo} and \eqref{eq:widebandslope-poistwostate-theo} are obtained by plugging \eqref{eq:dotRavg0MMPP} and \eqref{eq:ddotRavg0MMPP} into (\ref{eq:ebnomin-ra}) and \eqref{eq:widebandslope-ra}.
%
\hfill $\blacksquare$

\end{document}